\documentclass{IEEEtran}
\usepackage{graphicx}
\usepackage{amsmath, amssymb, bm}
\usepackage{subfigure}
\usepackage{epstopdf}
\usepackage{amsthm}
\usepackage{cite}
\usepackage{enumitem}
\usepackage{mathtools}
\usepackage{multirow}
\usepackage{bbm}
\usepackage{color}

\newtheorem{corollary}{Corollary}
\newtheorem{definition}{Definition}

\newcommand{\la}{\lambda}

\begin{document}
\setlist[description]{font=\normalfont}

\title{Spectral Domain Sampling of Graph Signals}
\author{Yuichi~Tanaka,~\IEEEmembership{Member,~IEEE}
\thanks{This work was supported in part by JST PRESTO under Grant JPMJPR1656.
}
\thanks{The author is with Graduate School of BASE, Tokyo University of Agriculture and Technology, Koganei, Tokyo, 184-8588 Japan, and also with PRESTO, Japan Science and Technology Agency, Kawaguchi, Saitama, 332-0012, Japan (email: ytnk@cc.tuat.ac.jp).}
\thanks{MATLAB code examples are available at http://tanaka.msp-lab.org/software.}
}

\markboth{IEEE Transactions on Signal Processing}{Tanaka: Spectral Domain Sampling of Graph Signals}
\maketitle

\begin{abstract}
Sampling methods for graph signals in the graph spectral domain are presented. Though conventional sampling of graph signals can be regarded as sampling in the graph vertex domain, it does not have the desired characteristics in regard to the graph spectral domain. With the proposed methods, the down- and upsampled graph signals inherit the frequency domain characteristics of the sampled signals defined in the time/spatial domain. The properties of the sampling effects were evaluated theoretically in comparison with those obtained with the conventional sampling method in the vertex domain. Various examples of signals on simple graphs enable precise understanding of the problem considered. Fractional sampling and Laplacian pyramid representation of graph signals are potential applications of these methods.
\end{abstract}


\section{Introduction}
\subsection{Motivation}
Sampling is a fundamental tool in discrete signal processing \cite{Oppenh2009, Rabine1978, Vaidya1993, Vetter2014}. It is used to convert the signal rate and is thus a key tool for multirate signal processing. For signals in the time or spatial domain, e.g., audio, speech, and image signals, the definition of sampling is intuitive; that is, for downsampling by two, every other sample is taken, and for upsampling, zeros are inserted between the samples.

Downsampling in the frequency domain broadens the signal bandwidth \cite{Vaidya1993, Vetter1995, Strang1996, Vetter2014}, thereby causing aliasing of non-bandlimited signals. On the other hand, upsampling narrows the bandwidth and creates imaging components. To avoid such aliasing and imaging, a low-pass filter is applied to signals before downsampling and after upsampling.

This intuitive definition of sampling also applies to graph signal processing \cite{Shuman2013, Ortega2018, Sandry2013}, a rapidly growing research area. Many potential applications of graph signal processing, for example, analyzing, restoring, and compressing signals on irregularly structured networks (such as data in social/transportation/neuronal networks, point cloud attributes, and multimedia signals), have been explored \cite{Leonar2013, Zhang2014, Ono2015, Hu2015, Miller2015, Onuki2016, Segarr2016, Higash2016, Pang2017, Shahid2016, Cheung2018, Yamamo2016, Liu2017, Perrau2017}.

A graph signal is defined as a discrete signal $\bm{f}\in \mathbb{R}^N$ for which the $n$th sample $f[n]$ is located on the $n$th vertex of a graph (graph and graph signal are formally defined in Section \ref{sec:notations}). In graph signal processing, downsampling means reducing the size of the graph as well as reducing the number of samples. Upsampling means expanding the graph as well as increasing the number of samples. Many approaches to reducing the size of graphs while attempting to retain at least some of the characteristics of the original graph have been proposed \cite{Narang2010, Ron2011, Narang2011, Narang2012, Dorfle2013, Nguyen2015, Shuman2016, Aspval1984}. However, few approaches to increasing the size of graphs have been proposed since doing so requires estimating the characteristics of the expanded graph \cite{Tanaka2014a, Sakiya2014a}.

Several techniques for reducing graph size in regards to signal processing and (spectral) graph theory have been proposed; however, sampling of a graph signal itself has been limited to an intuitive approach, namely, simply retaining the original signal values on the vertices that remain in the reduced-size graph. This approach is called ``vertex domain sampling" here, and it is intuitively related to its counterpart for time domain signals. However, as mentioned later, \textit{vertex domain sampling does not inherit the frequency domain properties of sampled signals in the time domain}. That is, the spectral shape of the signal is different after vertex domain sampling.

This change in spectral shape could be a problem when designing multirate systems for graph signals (like graph wavelets and filter banks) \cite{Hammon2011, Narang2012, Narang2013, Shuman2016a, Tanaka2014a, Shuman2015, Tay2015, Ekamba2015, Sakiya2016a, Teke2016a, Teke2016b, Trembl2016, Jin2017} and multiscale transforms \cite{Sakiya2014a, Shuman2016} since the analogy from classical signal processing cannot be used. Therefore, the sampling of graph signals has to be carefully considered and designed so as to satisfy the frequency domain requirements.

In this study, methods for sampling graph signals in the graph spectral domain are derived. The sampled graph signals naturally inherit the frequency characteristics of sampled time domain signals, e.g., increased (decreased) bandwidths. The properties of time, vertex, and spectral domain sampling are briefly summarized in Table \ref{tb:property}, and the details are presented in the following sections.

To intuitively and precisely understand the proposed sampling approach, the properties of graph signals sampled in the graph spectral domain were experimentally determined and compared with those sampled using the vertex domain method. Even for signals on simple graphs, i.e., path and grid graphs, vertex domain sampling did not retain the spectral shape after sampling while spectral domain sampling did. The proposed graph spectral sampling methods were applied to fractional sampling and Laplacian pyramid representation of graph signals, which are potential applications.

This paper is organized as follows. The following subsections clarify the contributions of this work, define the notation used, and present preliminaries concerning graph signal processing. Sampling methods for classical discrete signal processing are introduced in Section \ref{sec:samplingDSP}. Desired properties of sampling for graph signals and definitions of conventional vertex domain sampling are introduced in Section \ref{sec:vertexdomainsampling}. The proposed sampling methods are presented in Section \ref{sec:spectral_samp}. The effects of spectral domain sampling on signals on some graphs are illustrated in Section \ref{sec:examples}. A few potential applications of the proposed methods in comparison with existing methods are described in Section \ref{sec:applications}. Finally, the key points of the paper are summarized and future work is mentioned in Section \ref{sec:conclusion}.

\begin{table}[tp]
\caption{Properties of Sampled Signals. Details Are Described in Section \ref{subsec:desirableprop}.}
\centering
\begin{tabular}{l||p{.3in}|p{.3in}|p{.3in}|p{.3in}}
\hline
 & \multicolumn{2}{c|}{Discrete SP} & \multicolumn{2}{c}{Graph SP}\\\hline
Sampling domain & Time & Freq. & Vertex & \textbf{Spec.} \\\hline
Preservation of signal values& \checkmark & \checkmark & \checkmark & \\
in the time/vertex domain & & & \\\hline
Preservation of spectral shapes &  \checkmark & \checkmark &  & \checkmark\\
after sampling & & & \\\hline
\end{tabular}
\label{tb:property}
\end{table}%

\subsection{Related Work}
\subsubsection{Graph Signal Processing}
Downsampling of graph signals in the vertex domain has been extensively studied. The assumption in these studies was that the sampled signal is also a graph signal; that is, the sampled signal must have a corresponding reduced-size graph.

Several methods for reducing graph size have been proposed. They include graph coloring \cite{Narang2012, Narang2013, Harary1977, Aspval1984}, Kron reduction \cite{Shuman2016, Dorfle2013}, maximum spanning trees \cite{Nguyen2015}, weighted max-cut \cite{Narang2010}, and graph coarsening using algebraic distance \cite{Ron2011}. In these vertex domain methods, the vertices remaining after downsampling are selected, and they are reconnected as needed. Such methods usually need one-to-one mapping from the vertices of the reduced-size graph to those of the original graph. A method for multiscale graph reduction that does not necessarily need one-to-one mapping (but is still a vertex domain downsampling method) was proposed by Tremblay and Borgnat \cite{Trembl2016}. The relationship between the downsampling-then-upsampling operation in the vertex domain and the discrete Fourier transform (DFT) domain counterpart when the downsampling and upsampling are performed for a specific graph (called ``$\Omega$-structure" elsewhere \cite{Teke2016b}) was clarified by Teke and Vaidyanathan \cite{Teke2016b}. Note again, these methods are based on a vertex domain approach.

Pesenson considered the bandlimiting of graph signals followed by vertex domain sampling along with the study of sampling theory for graph signals \cite{Pesens2008}. However, the focus was still on vertex domain sampling: the down- and upsampling operators in the graph spectral domain were not fully defined. In contrast, there is no assumption here that a graph signal is bandlimited; instead, down- and upsampling, mimicking the frequency domain characteristics of discrete time domain signals, are considered. The relationship between the proposed method and the sampling theory of graph signals \cite{Chen2015, Anis2014, Wang2015, Anis2016, Tsitsv2016} is described in Section \ref{subsec:samplingtheorem}.

This work makes three main contributions:
\begin{itemize}
\item Vertex domain sampling is shown not to inherit the desirable spectral properties.
\item Methods for sampling graph signals defined on the graph spectral domain are presented.
\item The proposed sampling methods replicate the spectral domain characteristics that are expected as a counterpart of sampling in the frequency (i.e., DFT) domain.
\end{itemize}

\subsubsection{Other Related Work}
In this paper, ``sampling" means reducing or increasing the number of \textit{discrete} signal samples. This means that the original signal itself is still defined in the discrete domain and that no assumptions are made about the continuous domain counterparts of the graph signals. However, various studies can be found relating discrete signals and their continuous domain counterparts not necessarily lying in a Euclidean domain.

Without being exhaustive, P\"uschel and Moura introduced algebraic signal processing \cite{Pusche2008, Pusche2008a}, which is an ancestor of graph signal processing. Kova\v{c}evi\'c and P\"uschel generalized the sampling of 1-D and 2-D discrete/continuous signals with definitions of shift and boundary conditions that differ from those for ordinary time/spatial domain signals \cite{Kovace2010}. This sampling was further generalized to graph signal processing \cite{Sandry2013, Ortega2018, Shuman2013}, which can treat signals on arbitrary graphs.

There have also been studies on the relationship between the (continuous) Laplace--Beltrami operator and the graph Laplacian. For example, an approximation to the Laplace--Beltrami operator of a point cloud in $d$-dimensional space was proposed \cite{Belkin2009}. Convergence of the graph Laplacian to the Laplace--Beltrami operator has been studied \cite{Singer2006, Ting2010, Belkin2008, Hein2005}. Schemes for discretizing Laplace--Beltrami operators over triangulated surfaces have been proposed \cite{Xu2004}. Methods for estimating the graph Laplacian from Laplace--Beltrami operators have also been studied \cite{Gine2006}.

Note that the original graph signal considered in this paper may or may not have a counterpart continuous signal. For example, friendship networks, web structures, and molecule structures are not generally assumed to have continuous counterparts. Knowledge of a continuous domain signal (if one exists) could help in the development of efficient and reliable graph signal processing systems; however, such development remains an open problem.

\begin{figure*}
\centering
\subfigure[][Downsampling of time domain signals]{\includegraphics[width=.54\linewidth]{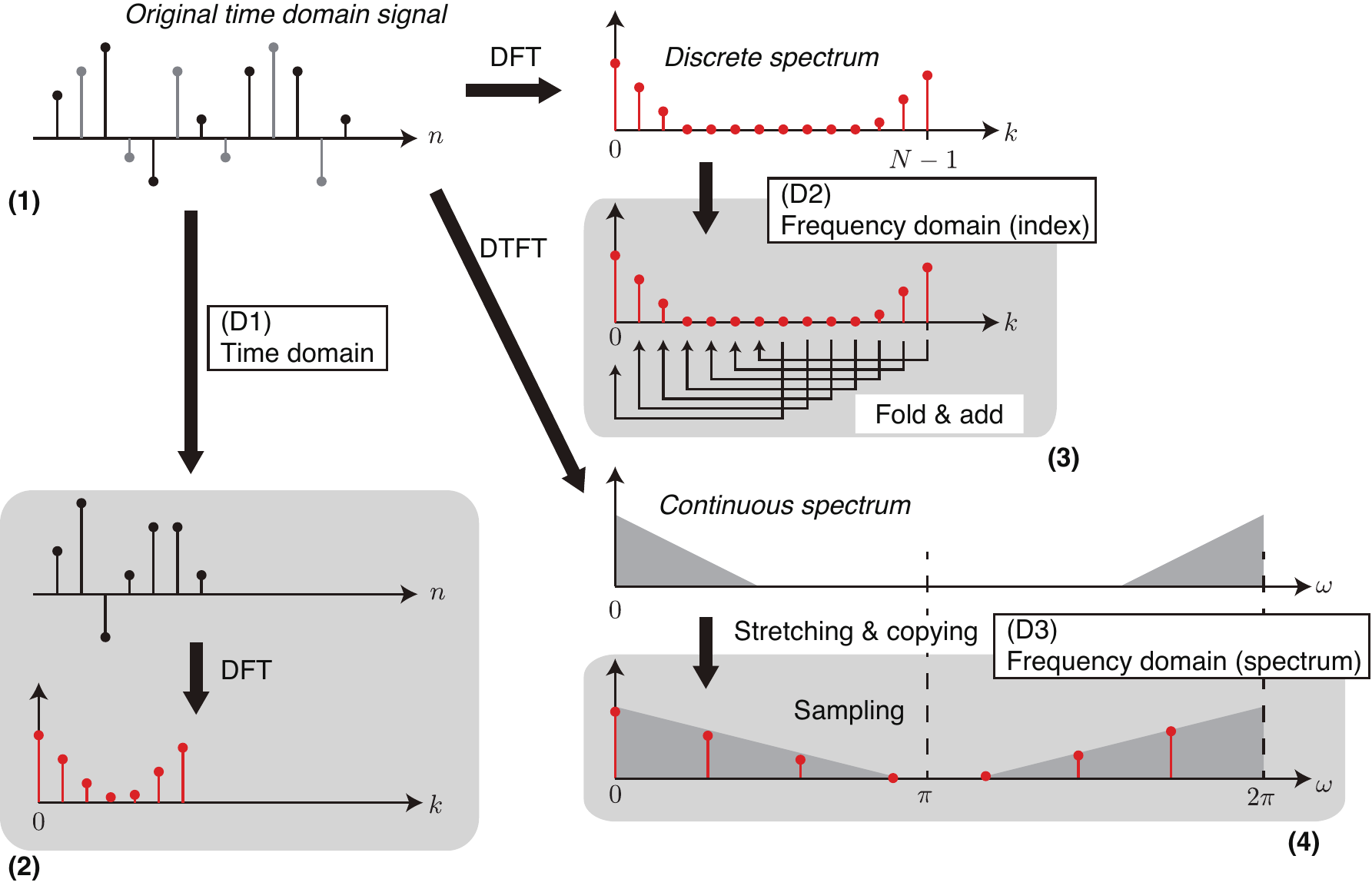}}\ 
\subfigure[][Upsampling of time domain signals]{\includegraphics[width=.44\linewidth]{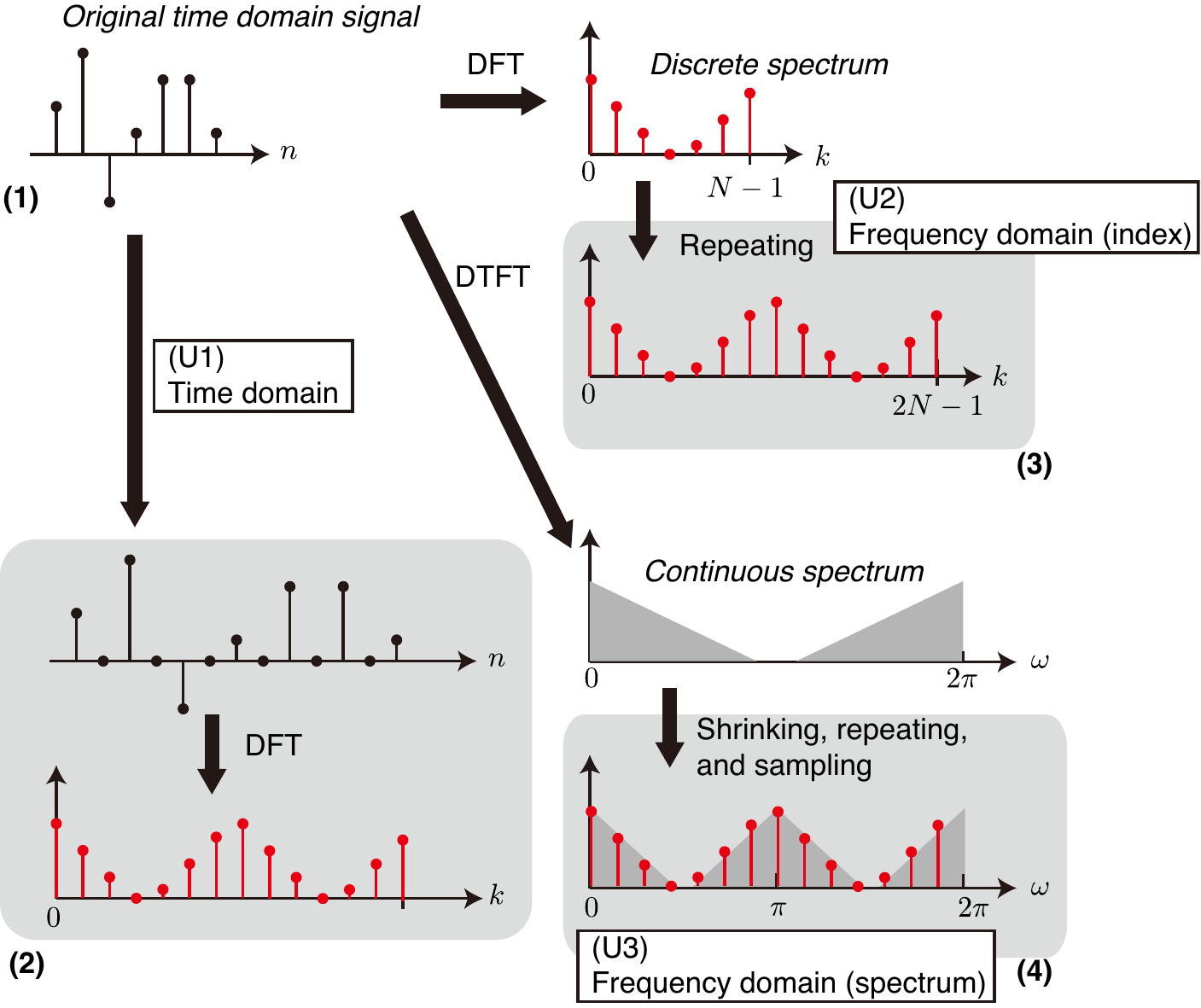}}
\caption{Sampling of discrete time domain signals. (a) Downsampling of time domain signals. Signal is downsampled by two and bandlimited. Shaded areas represent identical signals. (1) Original signal. Gray-colored samples are removed. (2) Downsampling method (D1): direct downsampling in the time domain. (3) Downsampling method (D2): frequency domain downsampling based on signal index. (4) Downsampling method (D3): frequency domain downsampling based on continuous spectrum. (b) Upsampling of time domain signals. Signal is upsampled by two. Shaded areas represent identical signals. (1) Original signal. (2) Upsampling method (U1): direct upsampling in time domain. (3) Upsampling method (U2): frequency domain upsampling based on signal index. (4) Upsampling method (U3): frequency domain upsampling based on continuous spectrum.}
\label{fig:sampling_timedomain}
\end{figure*}

\subsection{Notation}
\label{sec:notations}
A graph $\mathcal{G}$ is represented as $\mathcal{G} = (\mathcal{V}, \mathcal{E})$, where $\mathcal{V}$ and $\mathcal{E}$ denote sets of vertices and edges, respectively. The number of vertices is given as $N=|\mathcal{V}|$ unless otherwise specified. The $(m,n)$-th element of adjacency matrix $\mathbf{A}$ is $w_{mn}$. It represents the weight of the edge between the $m$th and $n$th vertices; $w_{mn} = 0$ for unconnected vertices. Degree matrix $\mathbf{D}$ is a diagonal matrix, and its $m$th diagonal element is $d_{mm} = \sum_n w_{mn}$.

Graph signal processing uses different variation operators \cite{Anis2016} depending on the application and assumed signal and/or network models. Hereafter, the variation operator considered is a combinatorial graph Laplacian for a finite undirected graph with no loops or multiple links $\mathbf{L}:=\mathbf{D}-\mathbf{A}$.

Key symbols are as follows:
\begin{enumerate}
\item $f: \mathcal{V}\rightarrow \mathbb{R}$: graph signal that assigns one value to each vertex. It can be written as a vector $\bm{f}$ in which the $n$th element $f[n]$ represents the signal value at the $n$th vertex.
\item $\bm{u}_{i}$: the $i$th eigenvector of $\mathbf{L}$.
\item $\lambda_i$: the $i$th eigenvalue of $\mathbf{L}$, i.e., $\mathbf{L}{\bm u}_{i} = \lambda_i {\bm u}_{i}$.
\end{enumerate}
Since $\mathbf{L}$ is a real symmetric matrix, $\mathbf{L}$ can always be decomposed into $\mathbf{L} = \mathbf{U} \bm{\Lambda} \mathbf{U}^*$, where $\mathbf{U} = [{\bm u}_{0}, \ldots, {\bm u}_{N-1}]$ is a unitary matrix, $\bm{\Lambda} = \text{diag}(\la_0, \la_1, \ldots, \la_{N-1})$, and $\cdot^*$ represents the conjugate transpose of a matrix. In this paper, $\la_i$ is often called \textit{graph frequency}.

Although the order of the eigenvalues is arbitrary, they are usually ordered as $0 = \la_0 < \la_1 \le \ldots \le \la_{N-1} = \la_{\max}$ for connected graphs. Hereafter, this order is used unless otherwise specified. There is freedom to choose the order for repeated eigenvalues. This was experimentally investigated (see Section \ref{subsec:repeatedev}).
 
The graph Fourier transform is defined as
\begin{equation}
\label{ }
\widetilde{f}[i] = \langle{\bm u}_{i}, \bm{f}\rangle = \sum_{n=0}^{N-1}u^*_{i}[n]f[n].
\end{equation}
Different definitions of graph Fourier transforms that treat repeated eigenvalues or irregularity of the eigenvalue distribution have also been proposed \cite{Deri2017, Giraul2018}.

\section{Sampling in Discrete Signal Processing}\label{sec:samplingDSP}
In classical (discrete) signal processing, sampling is intuitive. Down- and upsampling of one-dimensional signals in both the time and frequency domains are reviewed in the following subsections. For the sake of simplicity, $\bm{f}$ is assumed to be an ordinary time domain signal with length $N$.

\begin{figure*}[t]%
\centering
{\includegraphics[width=.7\linewidth]{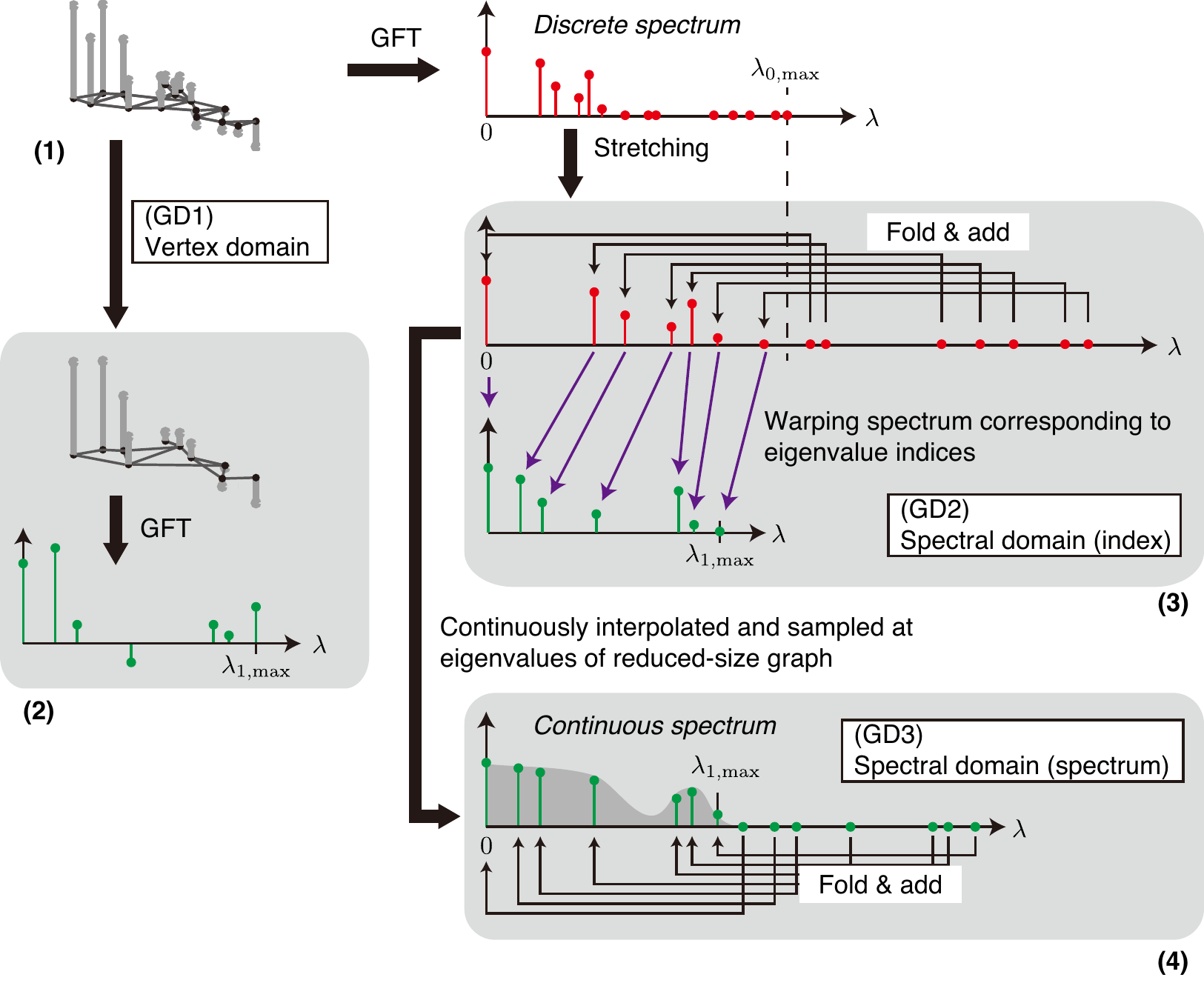}}
\caption{Downsampling of signals on graphs. Signal is downsampled by two and bandlimited. Shaded areas represent \textit{different} signals. (1) Original graph signal. (2) Downsampling method (GD1): direct downsampling in vertex domain. (3) Downsampling method (GD2): graph-spectral domain downsampling based on signal index. (4) Downsampling method (GD3): graph-spectral domain downsampling based on continuous spectrum.}
\label{fig:gd2and3}
\end{figure*}

\subsection{Downsampling}
First, downsampling is defined. For simplicity, it is assumed that the length of the original signal, $N$, is a multiple of the downsampling factor $M$. Let $F[k] := \sum_{n=0}^{N-1} f[n] e^{-j\frac{2\pi k}{N}n}$ be the DFT of time domain signal $\bm{f}$. The following sampling methods are illustrated in Fig. \ref{fig:sampling_timedomain}(a).

\begin{definition}[Downsampling]\label{def:ds_classical}
When a time domain signal $\bm{f} \in \mathbb{R}^N$ is downsampled to $\bm{f}_d \in \mathbb{R}^{N/M}$, the following statements \emph{(D1)--(D3)} are equivalent \cite{Oppenh2009, Rabine1978, Vaidya1993, Vetter2014}.
\end{definition}
\begin{description}
\item[(D1)] \textit{Time domain:} Retain every $M$th sample.
\begin{equation}
\label{ }
f_d[n] = f[Mn].
\end{equation}
 \item[(D2)] \textit{Frequency domain (index):} Retain every $M$th sample of the DFT spectrum $F[k]$ and sum aliasing terms.
\begin{equation}
\label{ }
F_d[k] = \sum_{p=0}^{M-1} F\left[\frac{pN}{M} + k\right],
\end{equation}
where $k = 0, \ldots, N/M$.
 \item[(D3)] \textit{Frequency domain (spectrum):} Sample stretched discrete-time Fourier transform (DTFT) spectrum $F(\omega) = \sum_{n=0}^{N-1} f[n]e^{-j\omega n}$.
\begin{equation}
\label{ }
F_d[k] = F_D\left(\frac{2\pi M}{N}k\right),
\end{equation}
where
\begin{equation}
\label{ }
F_D(\omega) = \frac{1}{M} \sum_{p=0}^{M-1}F\left(\frac{\omega - 2\pi p}{M}\right).
\end{equation}
\end{description}

\subsection{Upsampling}
The upsampling operator can be similarly defined and is illustrated in Fig. \ref{fig:sampling_timedomain}(b).

\begin{definition}[Upsampling]\label{def:us_classical}
The following statements, (U1)--(U3), concerning upsampling of time domain signal $\bm{f} \in \mathbb{R}^N$ by $L$ into $\bm{f}_u \in \mathbb{R}^{NL}$ are equivalent \cite{Vaidya1993}.
\end{definition}
\begin{description}
 \item[(U1)] \textit{Time domain:} Insert zeros.
\begin{equation}
\label{ }
f_u[n] = \begin{cases}
f[n/L] & \text{if } n = mL\\
0 & \text{otherwise.}
\end{cases}
\end{equation} 
 \item[(U2)] \textit{Frequency domain (index):} Repeat DFT spectrum $L$ times.
\begin{equation}
\label{ }
F_u[pN + k] = F[k],
\end{equation}
where $p = 0, 1, \ldots, L-1$.
 \item[(U3)] \textit{Frequency domain (spectrum):} Sample compressed DTFT spectrum $F(\omega)$.
\begin{equation}
\label{ }
F_u[pN + k] = F\left(\frac{2\pi}{N}k\right),
\end{equation}
where $p = 0, 1, \ldots, L-1$.
\end{description}

Definitions \ref{def:ds_classical} and \ref{def:us_classical} are always true in classical signal processing because the DFT spectrum, $F[k]$, is the version of $F(\omega)$ sampled at equal intervals $\frac{2\pi}{N}$; i.e., $F[k] = F\left(\frac{2\pi}{N}k\right)$.

\begin{figure*}[t]%
\centering
{\includegraphics[width=.7\linewidth]{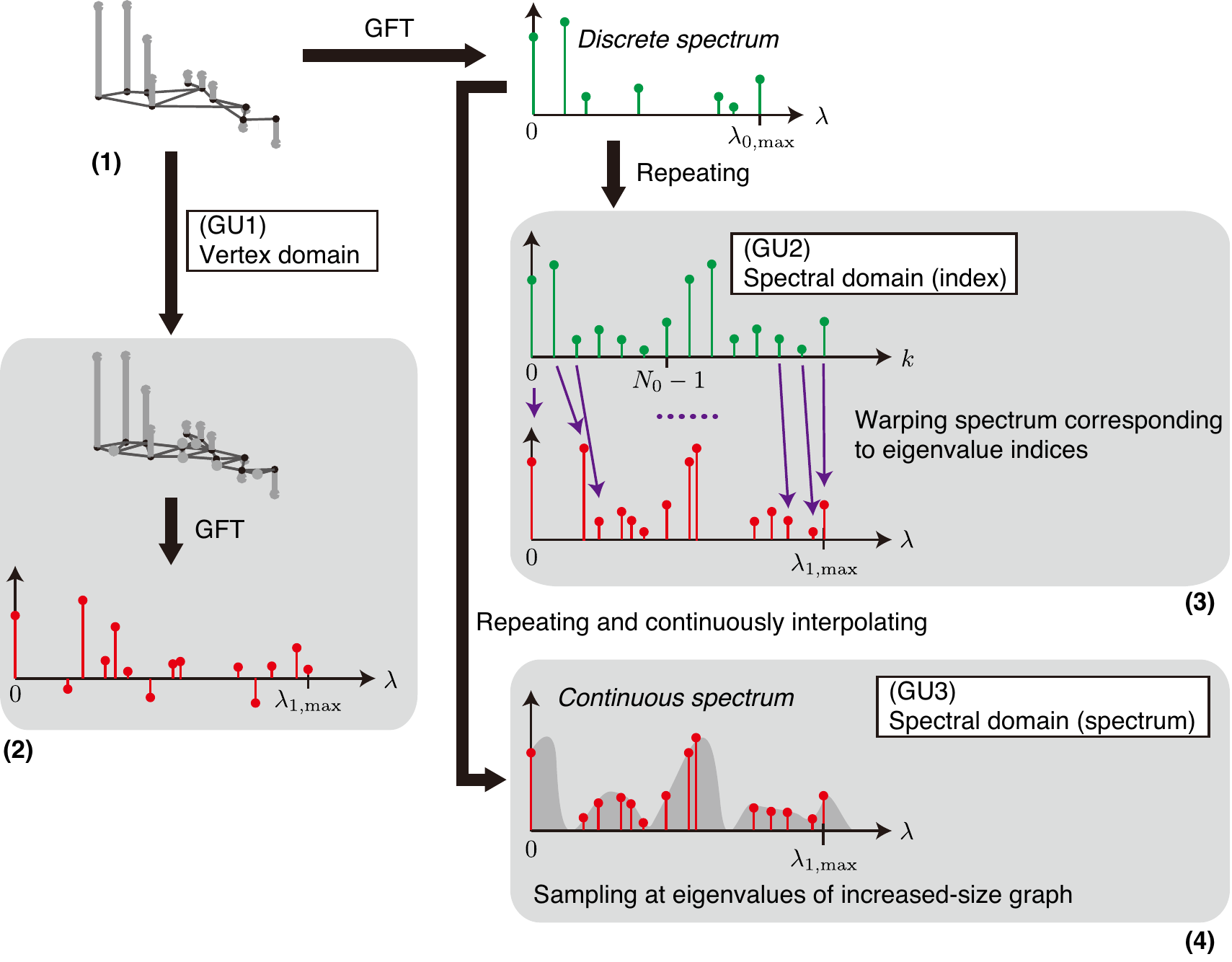}}
\caption{Upsampling of signals on graphs. Signal is upsampled by two. Shaded areas represent \textit{different} signals. (1) Original graph signal. (2) Upsampling method (GU1): direct upsampling in vertex domain. (3) Upsampling method (GU2): graph-spectral domain upsampling based on signal index. (4) Upsampling method (GU3): graph-spectral domain upsampling based on continuous spectrum.}
\label{fig:gu2and3}
\end{figure*}

\section{Sampling in Graph Signal Processing}\label{sec:vertexdomainsampling}
The sampling in graph signal processing is reviewed in this section. First, the desiderata of sampled graph signals are described for intuitively understanding the proposed approach. Next, the widely used vertex domain sampling method is explicitly defined. Finally, it is shown through several examples that vertex domain sampling does not always satisfy the desired behavior in the graph spectral domain.

\subsection{Desired Properties of Sampled Graph Signals}\label{subsec:desirableprop}
Let $M$ and $L$ be the down- and upsampling rates, respectively. In the design of multiscale and multirate graph signal processing systems, several characteristics of the sampled time domain signals should be satisfied after sampling of the graph signals, as shown in the following list of desiderata. These are the requirements for sampled graph signals; those for reduced- or increased-size graphs (not signals) are described elsewhere \cite{Shuman2016} and the references therein.

\begin{description}
 \item[(P1)] The sampled signal values in the vertex domain should remain unchanged.
 \item[(P2)] The width of the spectrum of the downsampled signal should be $M$ times broader than the original spectrum. Additionally, the aliasing components, i.e., the graph Fourier coefficients beyond the maximum graph frequency, should be \textit{folded} into the graph Fourier coefficients within the range $[0, \la_{\max}]$.\footnote{For variation operators other than the Laplacian, the frequency range is defined as $[\min(\text{Var}(\mathbf{L}, \bm{f})), \max(\text{Var}(\mathbf{L}, \bm{f}))]$, where $\mathbf{L}$ is a variation operator and $\text{Var}(\mathbf{L}, \bm{f})$ is a variation functional that measures the variation in $\bm{f}$ with respect to $\mathbf{L}$ \cite{Anis2016}.}
 \item[(P3)] The width of the spectrum of the upsampled signal should be $L$ times narrower than the original spectrum, and $L$ copies of the narrowed spectrum should be simultaneously obtained.
\end{description}

(P1) corresponds to time domain sampling. In (D1) and (U1), the original signal values are unchanged after sampling. (P2) corresponds to (D2) and (D3); the spectrum should be broader but retain the shape of the original spectrum if aliasing does not occur. If there is aliasing, the spectrum beyond the maximum frequency affects the broadened spectrum. (P3) corresponds to (U2) and (U3); upsampling shrinks the spectrum while retaining its shape, and copies created should be removed with a low-pass filter.

Although these properties are satisfied simultaneously when sampling discrete time domain signals, this is not the case when sampling graph signals. This is demonstrated in Section \ref{subsec:examples} after formal definitions of vertex domain sampling are given in the next subsection.

\subsection{Vertex Domain Sampling of Graph Signals}
The conventional and widely used method for sampling graph signals in the vertex domain, which corresponds to the intuitive counterpart of (D1) and (U1) in classical signal processing, is described as follows:

\begin{definition}[Downsampling of graph signals in vertex domain]\label{def:gd} Let $\mathcal{G}_0$ and $\mathcal{G}_1$ be the original graph and the reduced-size graph, respectively, where every vertex in $\mathcal{G}_1$ is in one-to-one correspondence with a vertex in $\mathcal{G}_0$. The original signal is $\bm{f}\in \mathbb{R}^N$. In the vertex domain, downsampling of $\bm{f}$ to $\bm{f}_d\in \mathbb{R}^{|\mathcal{V}_1|}$ is defined as follows.
\end{definition}
\begin{description}
  \item[(GD1)] \textit{Vertex domain:} Retain samples in $\mathcal{V}_1$.
\begin{equation}
\label{ }
f_{d}[n] = f[n'] \text{ if } v_{n'} \in \mathcal{V}_0 \text{ corresponds to } v_n \in \mathcal{V}_1.
\end{equation}
\end{description}
\begin{definition}[Upsampling of graph signals in vertex domain] $\mathcal{G}_0$ and $\mathcal{G}_1$ are the same as in Definition \ref{def:gd}. The original signal at this time is $\bm{f}\in \mathbb{R}^{|\mathcal{V}_1|}$ and its samples are associated with $\mathcal{G}_1$. Upsampling in the vertex domain, i.e., mapping from $\bm{f}$ to $\bm{f}_u\in \mathbb{R}^N$, is defined as follows.
\end{definition}
\begin{description}
  \item[(GU1)] \textit{Vertex domain:} Placing samples on $\mathcal{V}_1$ into the corresponding vertices in $\mathcal{G}_0$.
\begin{equation}
\label{}
f_{u}[n] = \begin{cases}
f[n'] & \text{if } v_{n'} \in \mathcal{V}_1 \text{ corresponds to } v_n \in \mathcal{V}_0\\
0 & \text{otherwise}.
\end{cases}
\end{equation}
\end{description}

They are illustrated in Figs. \ref{fig:gd2and3}-(2) and \ref{fig:gu2and3}-(2). It is clear that (GD1) and (GU1) satisfy (P1) by definition, but it is not known or has not been analyzed whether (P2) and (P3) are satisfied in general. Some counterexamples are shown next.

\subsection{Counterexamples}\label{subsec:examples}
Problems specific to graph signal processing are described using a few intuitive counterexamples as follows.

Let us start with the classical case in which it is assumed that a time domain signal is bandlimited: $\bm{f}$ has zero response at $\omega > |\pi/M|$. After downsampling by $M$, the spectrum of $\bm{f}_d$ is stretched by $M$. If the signal is \textit{not} bandlimited, aliasing occurs. Since downsampling methods (D1), (D2), and (D3) are equivalent, they produce the same stretching effect.

In contrast, this equivalence fails for vertex domain sampling of graph signals. The spectra of the graph signal sampled using (GD1) and (GU1) are represented as $\widetilde{\bm{f}}_{d} = \mathbf{U}^*_1\bm{f}_{d}$ and $\widetilde{\bm{f}}_{u} = \mathbf{U}^*_0\bm{f}_{u}$. The $\widetilde{\bm{f}}_{d}$ ($\widetilde{\bm{f}}_{u}$) depends on the original and reduced-(increased-)size graph Laplacian and the method used for graph reduction (expansion). 

Two simple cases are considered: ring and path graphs\footnote{In this example, the eigenvectors of the ring graph are real-valued vectors.}. It is assumed that $N$ is even and that the downsampling/upsampling ratio is $2$. It is also assumed that the size of the graph is reduced (or increased) in the most natural way: every other vertex is taken for downsampling, and a vertex is placed between every pair of vertices for upsampling.

\begin{figure}[tp]%
\centering
\subfigure[][Downsampling]{\includegraphics[width=.9\linewidth]{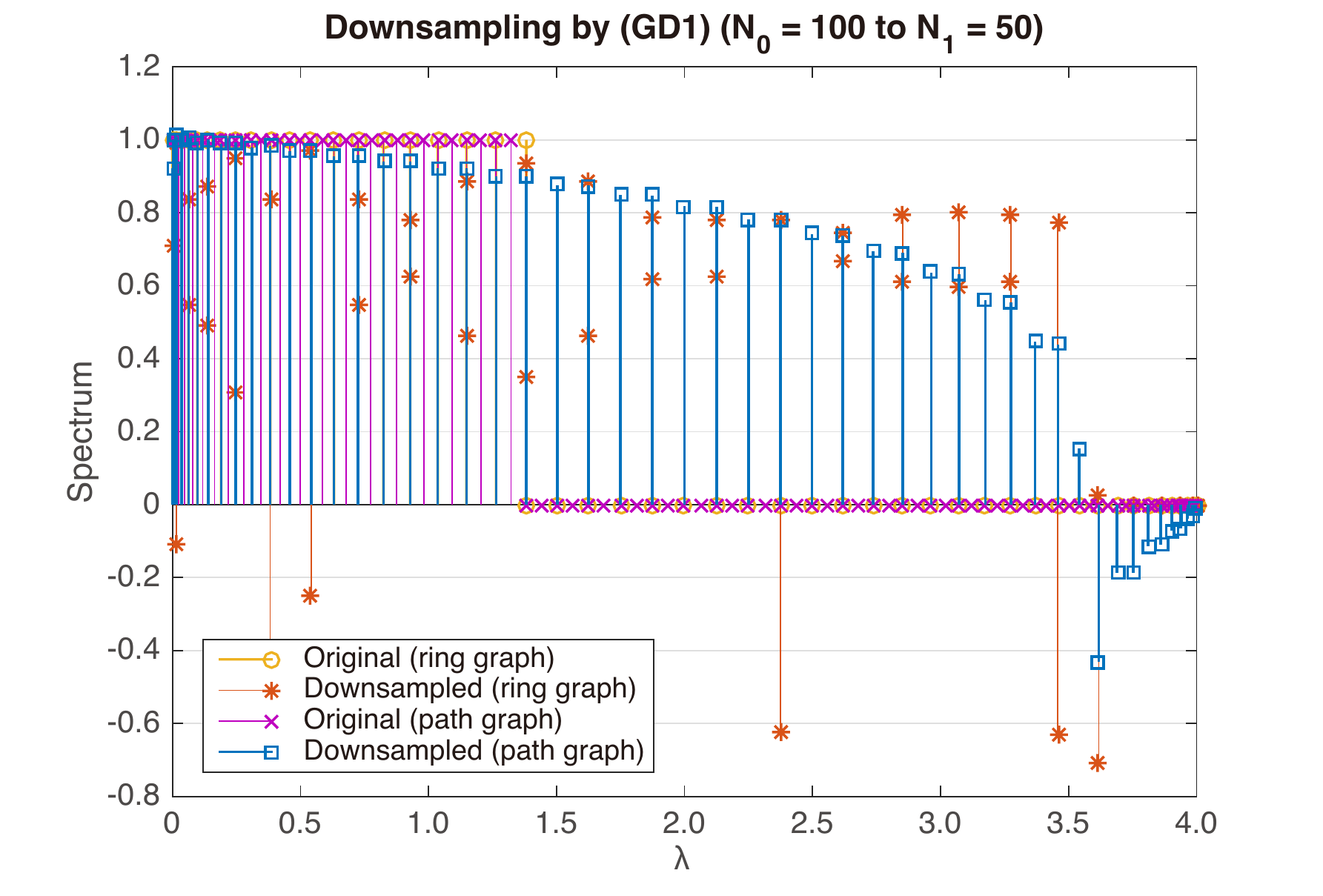}}\\
\subfigure[][Upsampling]{\includegraphics[width=.9\linewidth]{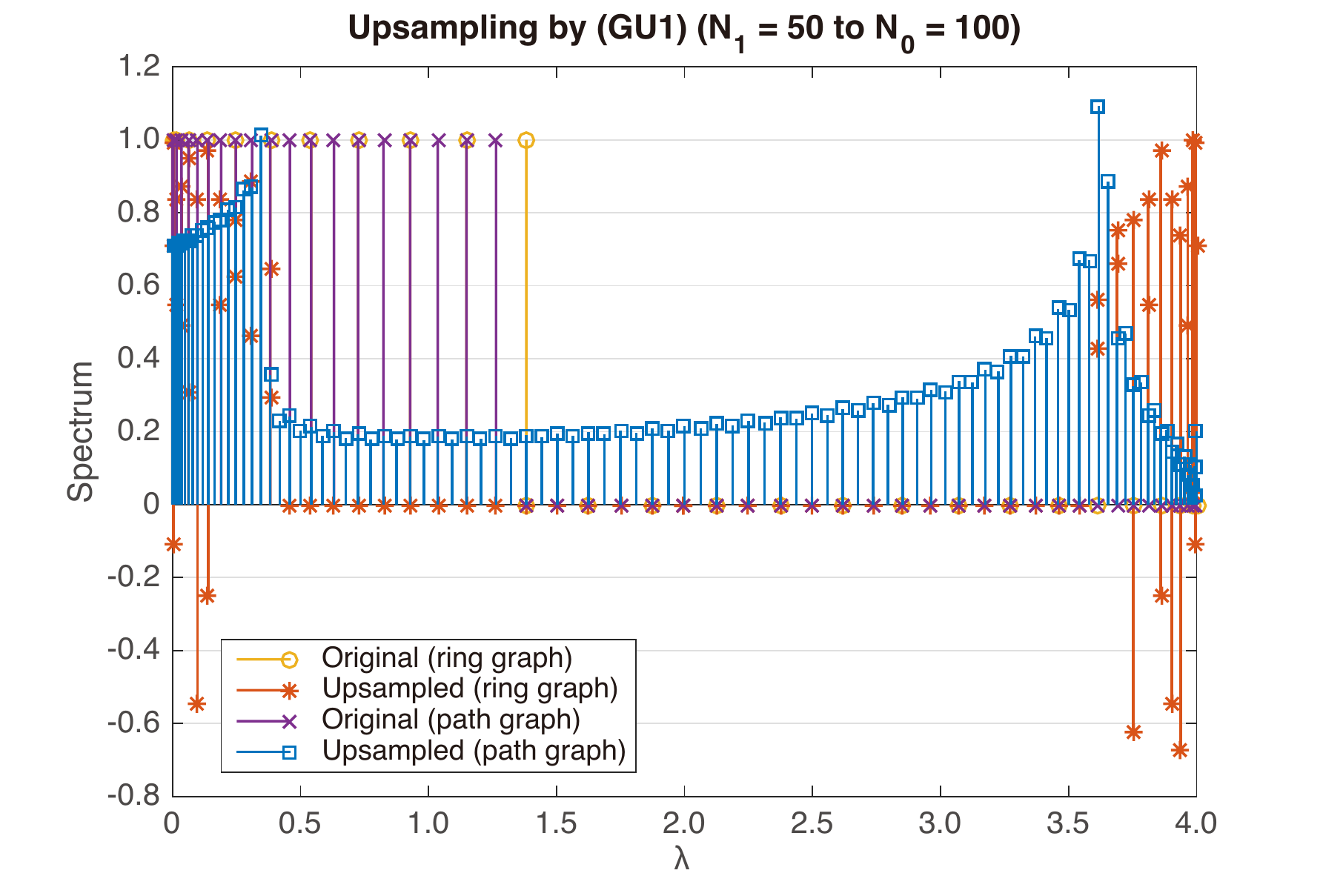}}
\caption{Spectral properties of sampled graph signals in vertex domain. For ring graph, stems are overlapped due to repeated eigenvalues.}
\label{fig:ex_gd1gu1}
\end{figure}

The spectra $\widetilde{\bm{f}}_{d}$ sampled using (GD1) and the spectra $\widetilde{\bm{f}}_{u}$ sampled using (GU1) are shown in Figs. \ref{fig:ex_gd1gu1}(a) and (b), respectively. Clearly, the characteristics in each graph spectral domain are \textit{not} as expected: the spectra are not stretched (or shrunk) and are completely different from the frequency domain counterpart.

Solving this problem was the motivation for this study. The approach taken was to define sampling operators of graph signals in the spectral domain.

\section{Spectral Domain Sampling}\label{sec:spectral_samp}
Methods respectively for down- and upsampling of graph signals in the graph spectral domain were derived, and their characteristics are described here. Since the spectrum of graph signals is generally one-sided, i.e., it can be defined in only the positive $\lambda$ \footnote{The eigenvalues can be both positive and negative for the other variation operators, but the variation functional representing graph frequency \cite{Anis2016} is still positive.}, how to stretch or shrink the spectrum has a certain degree of freedom. First, down- and upsampling in the spectral domain (which are closely related to DFT domain sampling in classical signal processing) are introduced, and then their slightly modified versions are explained\footnote{The sampling methods introduced in this paper can be extended to other variation operators like those introduced by Anis et al. \cite{Anis2016}, as long as the eigenvector matrix $\mathbf{U} \in \mathbb{C}^{N\times N}$ is invertible and the variation functionals associated with the variation operators \cite{Anis2016} can be defined appropriately, i.e., the variations in signals can be measured appropriately and ordered uniquely for a given $\bm{f}$ and the variation operator used. The behaviors of the sampled signals, in general, vary depending on the variation operator. Hence, spectral domain sampling for other operators needs further study.}.

\subsection{Downsampling}

\begin{definition}[Downsampling of graph signals in graph spectral domain]\label{def:GD_spectral} 
Let $\mathbf{L}_0 \in \mathbb{R}^{N \times N}$ and $\mathbf{L}_1 \in \mathbb{R}^{N/M \times N/M}$ be the graph Laplacians for the original graph and the reduced-size graph\footnote{$M$ is assumed to be divisor of $N$ for simplicity.}, respectively, and assume that their eigendecompositions are given by $\mathbf{L}_0 = \mathbf{U}_0 \bm{\Lambda}_0 \mathbf{U}^*_0$ and $\mathbf{L}_1 = \mathbf{U}_1 \bm{\Lambda}_1 \mathbf{U}^*_1$, where $\bm{\Lambda}_i = \textnormal{diag}(\la_{i,0}, \la_{i,1}, \ldots, \la_{i,\max})$. The downsampled graph signal $\bm{f}_{d} \in \mathbb{R}^{N/M}$ in the graph spectral domain is defined as follows.
\end{definition}

\begin{description}
 \item[(GD2)] \textit{Graph spectral domain (index):} The spectral response is divided by $M$, and the set of $N/M$ coefficients are summed.
\begin{equation}
\label{ }
\widetilde{f}_{d}[k] = \sum_{p=0}^{M-1} \widetilde{f}\left[\frac{pN}{M} + k\right].
\end{equation}
In matrix form, the downsampled graph signal can easily be represented as
\begin{equation}
\label{ }
\bm{f}_{d} = \mathbf{U}_1 \mathbf{S}_d \mathbf{U}^*_0 \bm{f},
\end{equation}
where $\mathbf{S}_d = \begin{bmatrix} \mathbf{I}_{N/M} & \mathbf{I}_{N/M} & \ldots \end{bmatrix}$. This downsampling method is illustrated in Fig. \ref{fig:gd2and3}-(3).

 \item[(GD3)] \textit{Graph spectral domain (spectrum):} The original discrete spectrum is stretched by $M$ and continuously interpolated. It is then sampled in accordance with the eigenvalue distribution of $\mathbf{L}_1$ and summed.
\begin{equation}
\label{eqn:fint}
\widetilde{f}_{d}[k] = \sum_{p=0}^{M-1} \widetilde{f}_{\text{int}}\left(\frac{\rho}{M}(\la_{1,k} + p \la_{1, \max})\right),
\end{equation}
where $\widetilde{f}_{\text{int}}(\la)$ ($\la \in [0, \la_{0,\max}]$) is the continuously interpolated version of $\widetilde{f}[k]$ and $\rho =\la_{0, \max}/\la_{1,\max}$.
\end{description}
This downsampling method is illustrated in Fig. \ref{fig:gd2and3}-(4). It is clear that (GD2) and (GD3) are the counterparts of (D2) and (D3) in classical signal processing.

As shown in Section \ref{subsec:interconnection}, the above definition of (GD2) has an immediate relationship to those of (D2) and (GD1) and is not problematic as long as the signal is bandlimited. However, even if slight aliasing occurs, (GD2) and (GD3) affect the low graph frequency components (discussed in Section \ref{subsec:aliasing}). This is due to the one-sided spectrum nature of $\widetilde{\bm{f}}$. In contrast, in classical signal processing, slight aliasing affects only the high-frequency components. Therefore, to maintain the conceptual relationship with classical frequency domain sampling, (GD2) and (GD3) are defined slightly differently:

\begin{description}
 \item[(GD2')] $\widetilde{\bm{f}}$ is evenly divided by $M$. The odd-numbered portions are then flipped and summed.
\begin{equation}
\label{ }
\begin{split}
\widetilde{f}_{d}[k] =& \sum_{p=0}^{\lceil M/2 \rceil -1} \widetilde{f}\left[\frac{2pN}{M} + k\right]\\
& + \sum_{p=0}^{\lfloor M/2 \rfloor -1} \widetilde{f}\left[\frac{2(p+1)N}{M} - k -1\right].
\end{split}
\end{equation}
This equation is easily represented in matrix form:
\begin{equation}
\label{ }
\bm{f}_{d} = \mathbf{U}_1 \mathbf{S}'_d \mathbf{U}^*_0 \bm{f},
\end{equation}
where $\mathbf{S}'_d = \begin{bmatrix} \mathbf{I}_{N/M} & \mathbf{J}_{N/M} & \mathbf{I}_{N/M} & \ldots \end{bmatrix}$, in which $\mathbf{J}$ is the counter-identity matrix.

 \item[(GD3')] The continuously interpolated spectrum is sampled symmetrically in accordance with the eigenvalue distribution of $\mathbf{L}_1$ and summed.
\begin{equation}
\label{ }
\begin{split}
\widetilde{f}_{d}[k] =& \sum_{p=0}^{\lceil M/2 \rceil -1} \widetilde{f}_{\text{int}}\left(\frac{\rho}{M}(\la_{1,k} + 2p \la_{1, \max})\right)\\
& + \sum_{p=0}^{\lfloor M/2 \rfloor -1} \widetilde{f}_{\text{int}}\left(\frac{\rho}{M}(-\la_{1,k} + 2(p+1) \la_{1, \max})\right).
\end{split}
\end{equation}
\end{description}

\subsection{Upsampling}
\begin{definition}[Upsampling of graph signals in the graph spectral domain]\label{def:GU_spectral} Let $\mathbf{L}_0 \in \mathbb{R}^{N \times N}$ and $\mathbf{L}_1 \in \mathbb{R}^{NL \times NL}$ be the graph Laplacians for the original graph and for the increased-size graph, respectively. The upsampled graph signal $\bm{f}_{u} \in \mathbb{R}^{NL}$ in the graph spectral domain is defined as follows.
\end{definition}

\begin{description}
 \item[(GU2)] \textit{Graph spectral domain (index):} The original spectrum is repeated $L$ times.
\begin{equation}
\label{ }
\widetilde{f}_{u}[pN+k] = \widetilde{f}[k],\ p= 0, \ldots, L-1 
\end{equation}
 \item[(GU3)] \textit{Graph spectral domain (spectrum):} The original spectrum is repeated $L$ times and continuously interpolated. The interpolated spectrum is then sampled in accordance with the eigenvalue distribution of $\la_1$.
\begin{equation}
\label{ }
\widetilde{f}_{u}[pN + k] = \widetilde{\underline{f}}_{\text{int}}\left(\rho L\la_{1,k}\right),\ p= 0, \ldots, L-1
\end{equation}
where $\widetilde{\underline{f}}_{\text{int}}(\la)$ ($\la \in [0, L \la_{0,\max}]$) is the interpolated version of the repeated spectrum $\widetilde{\underline{\bm{f}}} = [\underbrace{\widetilde{\bm{f}}^{\top}\quad \widetilde{\bm{f}}^{\top}\quad \cdots}_L]^\top
$.
\end{description}
These upsampling methods are illustrated in Fig. \ref{fig:gu2and3}; they are the counterparts of (U2) and (U3).

In the same manner as for downsampling, modified versions of (GU2) and (GU3) can be defined.
\begin{description}
 \item[(GU2')] The original and flipped spectra are alternatively repeated.
\begin{equation}
\label{ }
\widetilde{f}_{u}[pN+k] = \begin{cases}
\widetilde{f}[k] & p= 0, 2, \ldots, \lceil L/2 \rceil-1\\
\widetilde{f}[N - k - 1] & p= 1, 3, \ldots, \lfloor L/2 \rfloor-1.
\end{cases}
\end{equation}
 \item[(GU3')] The original and flipped spectra are alternatively repeated $L$ times and continuously interpolated. The interpolated spectra are then sampled in accordance with the eigenvalue distribution of $\la_1$.
\begin{equation}
\label{ }
\widetilde{f}_{u}[pN + k] = \widetilde{\underline{f}}'_{\text{int}}\left(\rho L\la_{1,k}\right),\ p= 0, \ldots, L-1
\end{equation}
where $\widetilde{\underline{f}}'_{\text{int}}(\la)$ is the interpolated version of $\widetilde{\underline{\bm{f}}}' = [\underbrace{\widetilde{\bm{f}}^{\top}\quad \widetilde{\bm{f}}^{\top}\mathbf{J}_N\quad \cdots}_{L}]^\top$.
\end{description}

\subsection{Avoiding Aliasing and Imaging}
Definitions \ref{def:GD_spectral} and \ref{def:GU_spectral} mean that aliasing (due to downsampling) and imaging (due to upsampling) can be avoided by using appropriate low-pass graph filters. For (GD2) and (GU2), the ideal filter characteristic in the graph spectral domain is defined as
\begin{equation}
\label{eqn:antialiasingfilter}
H[k] = \begin{cases}
1 & \text{if } k \leq \frac{N}{M} \text{ (GD2) or } k \leq N \text{ (GU2)},\\
0 & \text{otherwise}.
\end{cases}
\end{equation}
In contrast, for (GD3) and (GU3), the ideal filter characteristic is defined by the width of the spectrum:
\begin{equation}
\label{ }
H(\la) = \begin{cases}
1 & \text{if } \la \leq \frac{\la_{1, \max}}{M} \text{ (GD3) or } \la \leq \frac{\la_{1, \max}}{L} \text{ (GU3)},\\
0 & \text{otherwise}.
\end{cases}
\end{equation}
Use of these low-pass filters also prevent aliasing and imaging when (GD2'), (GD3'), (GU2'), and (GU3') are used instead of their original versions. Since (GD1) and (GU1) do not have an intuitive relationship between the main frequency components and the aliasing or imaging ones, it is generally difficult to prevent aliasing and imaging by using a low-pass filter.

\begin{figure*}[tp]%
\centering
\subfigure[][Downsampling: path graph]{\includegraphics[width=.45\linewidth]{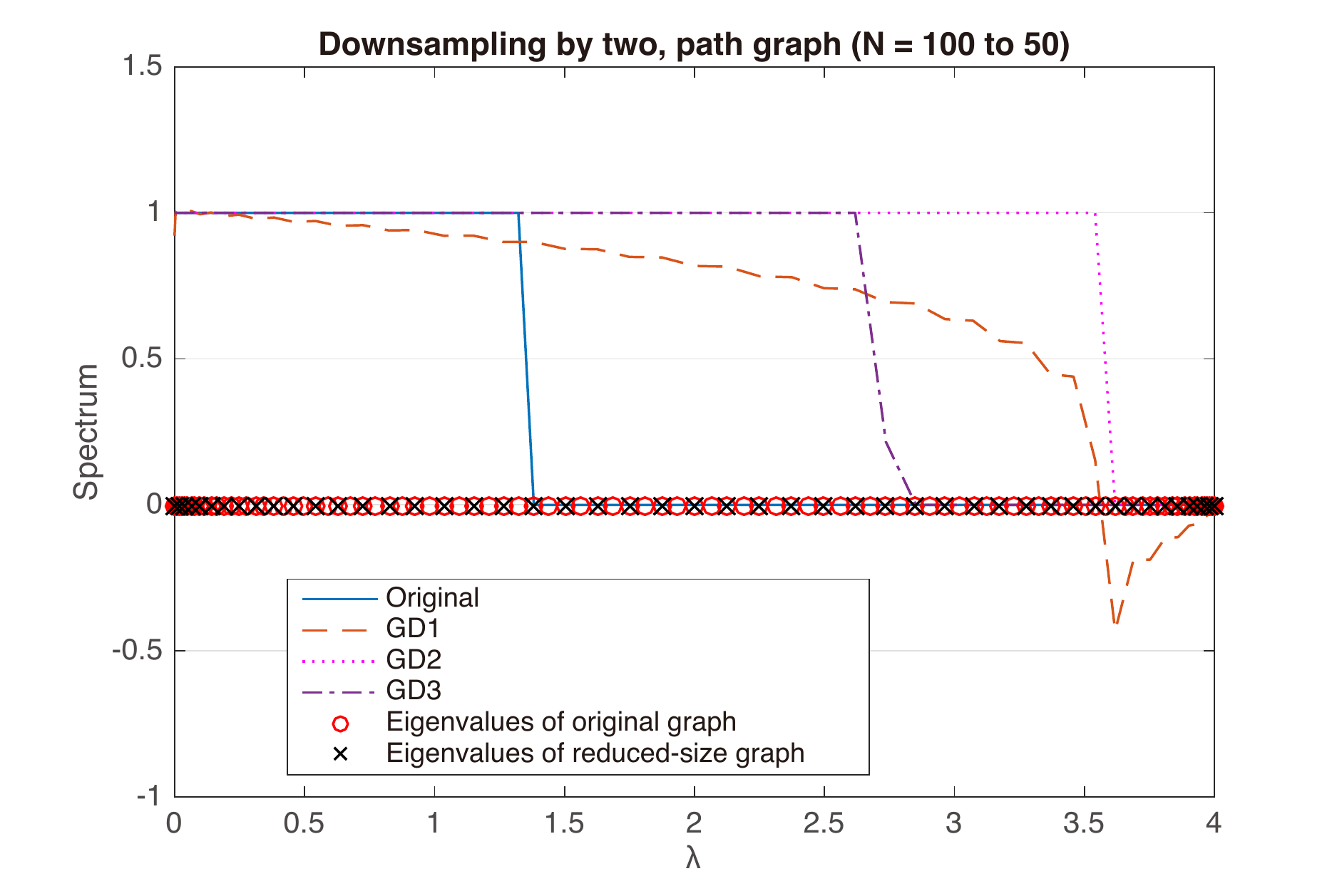}}\ 
\subfigure[][Upsampling: path graph]{\includegraphics[width=.45\linewidth]{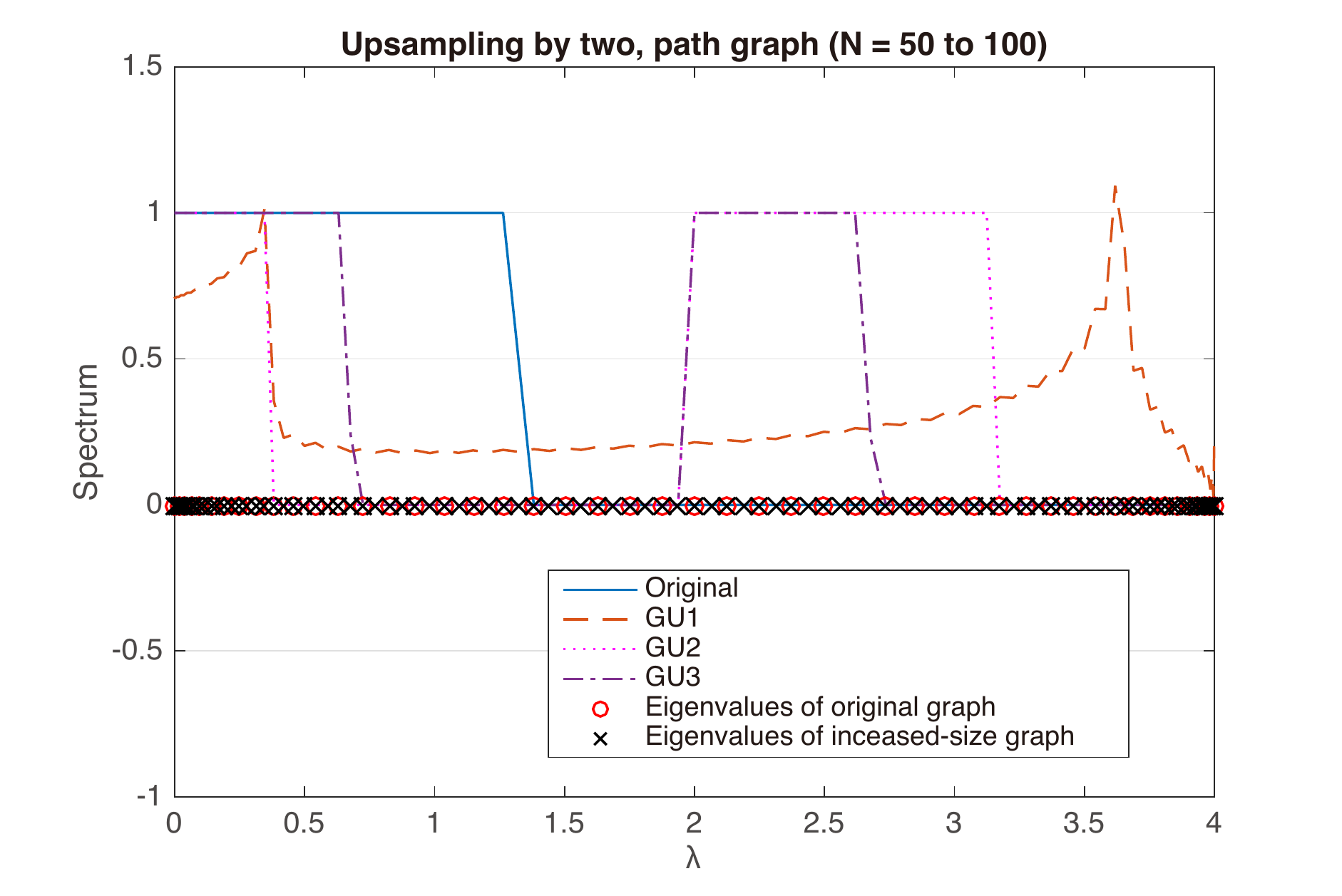}}\\
\subfigure[][Downsampling: grid graph]{\includegraphics[width=.45\linewidth]{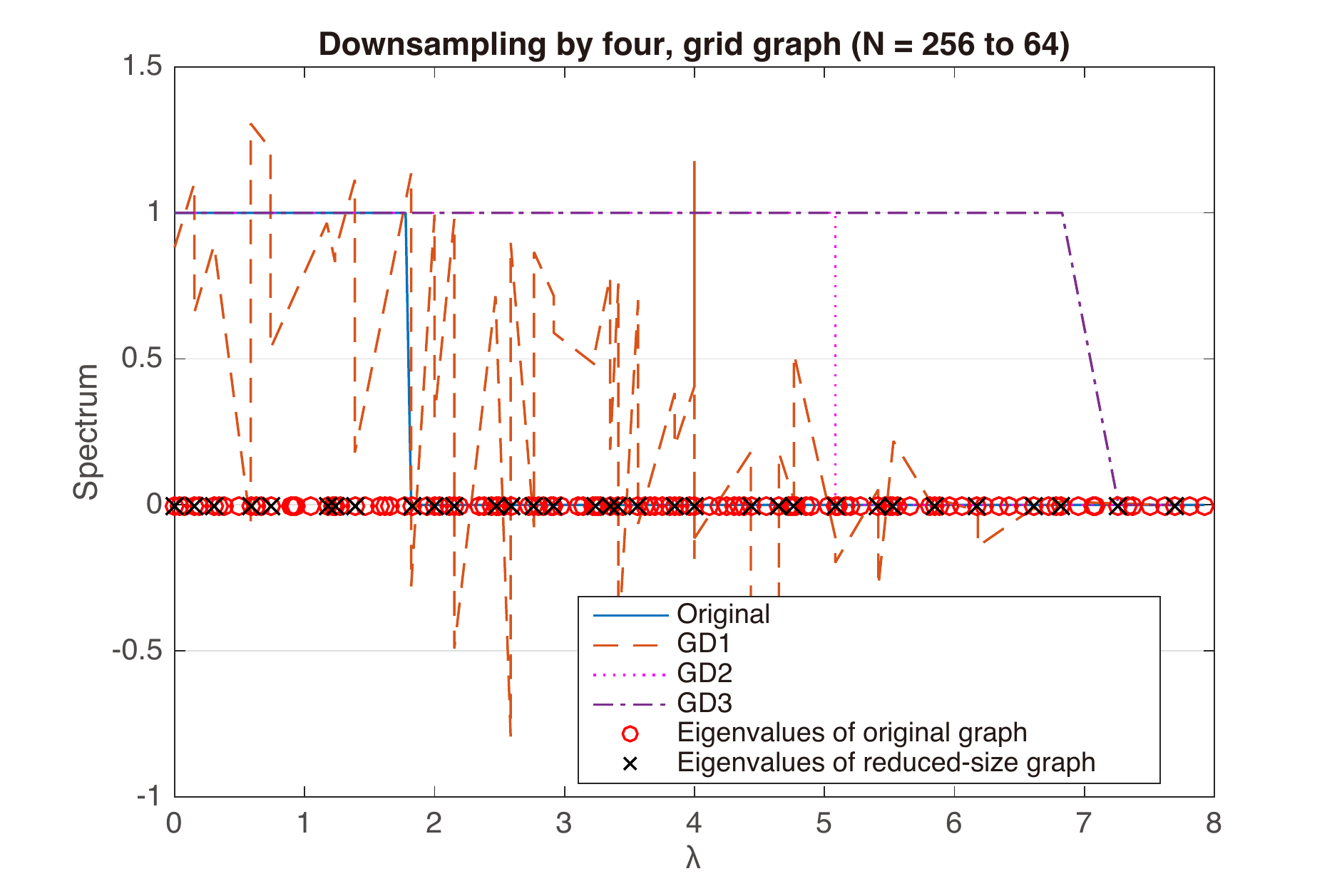}}\ 
\subfigure[][Downsampling: random regular graph]{\includegraphics[width=.45\linewidth]{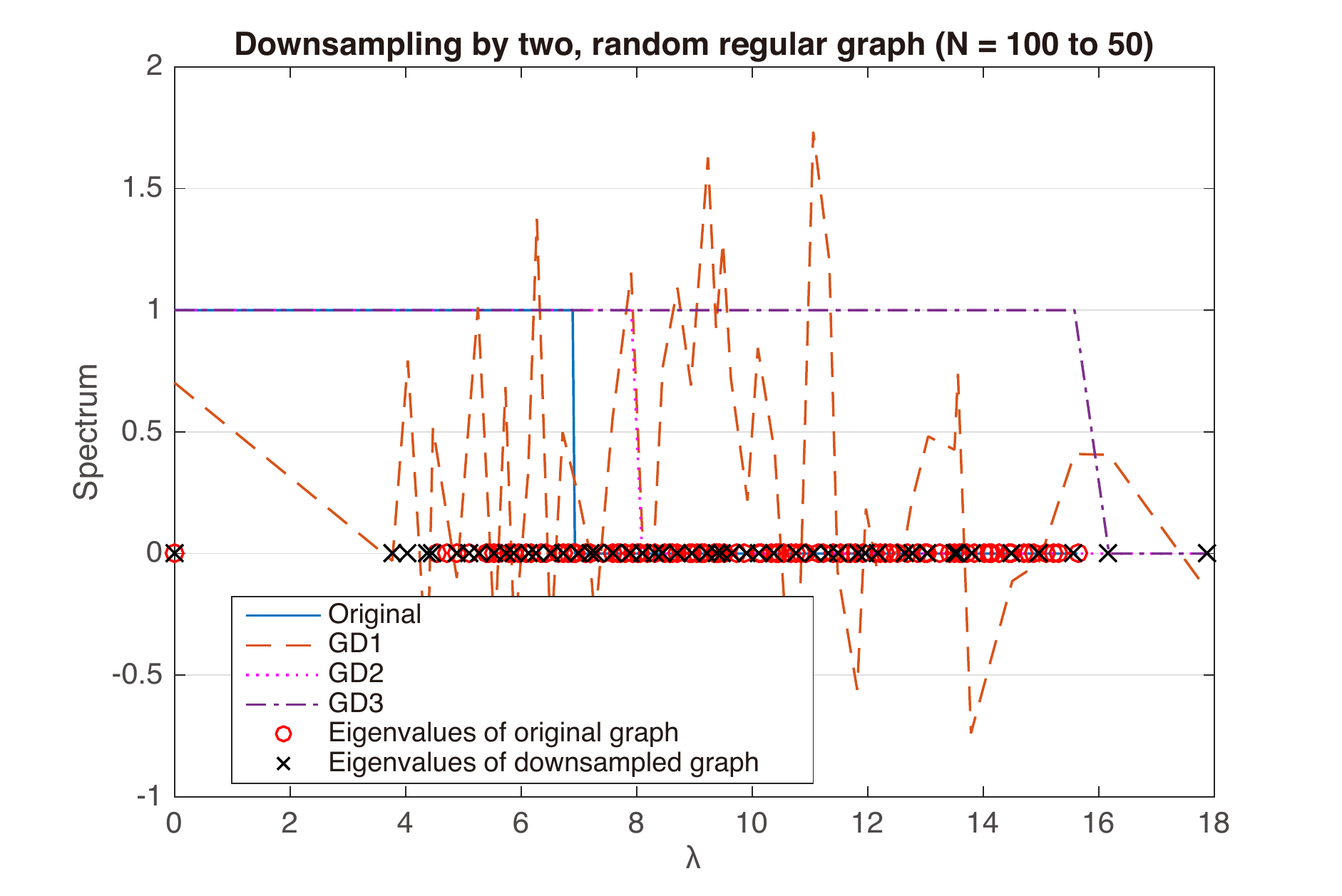}}
\caption{Illustrative examples of graph signal sampling.}
\label{fig:examples_downsampling}
\end{figure*}

\begin{figure}[t]%
\centering
{\includegraphics[width=.9\linewidth]{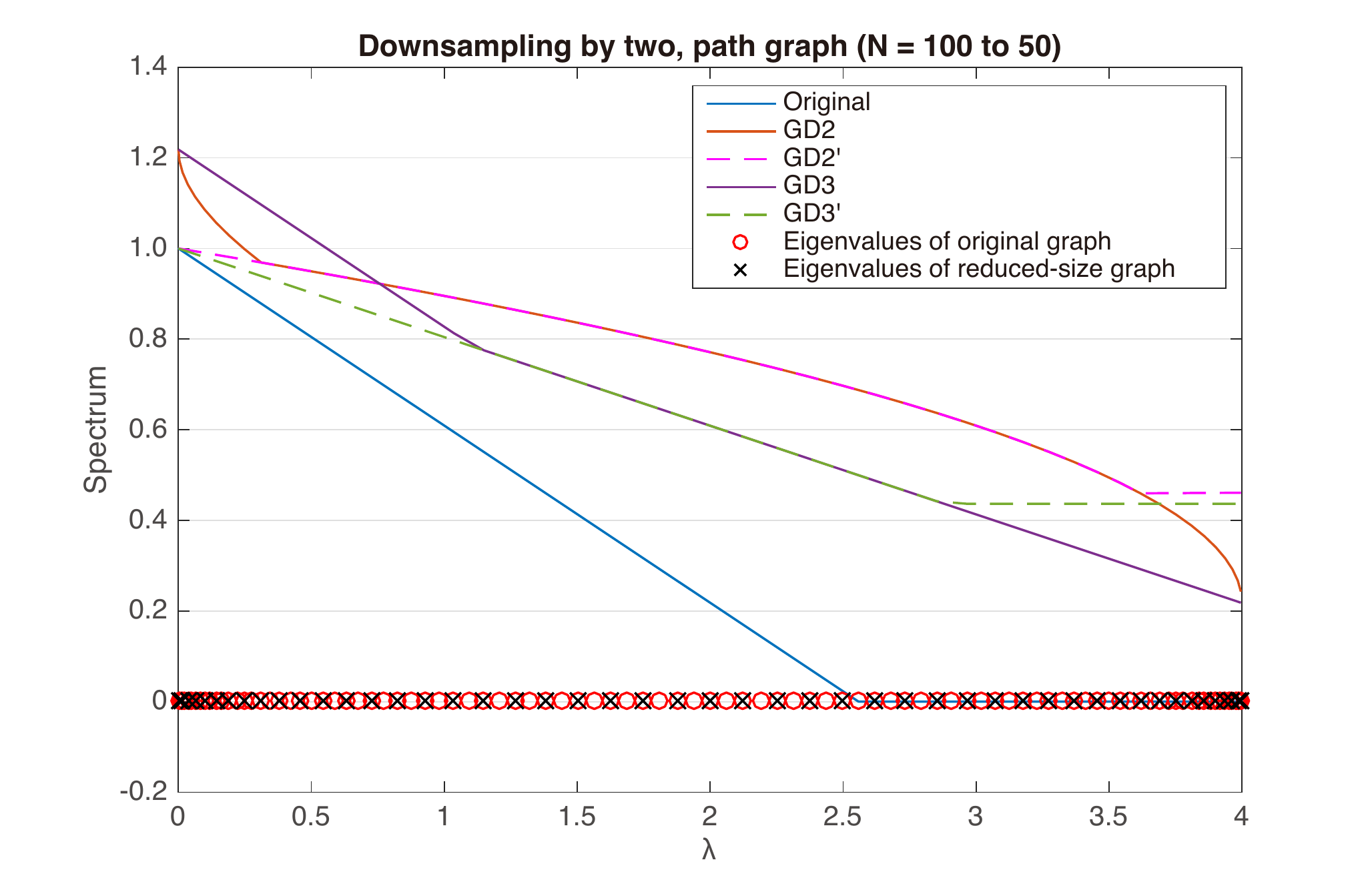}}
\caption{Illustrative examples of aliasing effects.}
\label{fig:path_aliasing}
\end{figure}

\subsection{Interconnections between Sampling Approaches}\label{subsec:interconnection}
From the definitions of (D2) and (GD2), the following relationship can be easily obtained.

\begin{corollary}\label{col:downsampling}
Assume that $\mathcal{G}_0$ and $\mathcal{G}_1$ are ring graphs with $N$ and $N/M$ vertices, respectively, where $N$ is a multiple of $M$. Additionally assume that $\mathbf{U}^*_i$ ($i\in\{0,1\}$) is a DFT matrix\footnote{The DFT matrix diagonalizes the graph Laplacian of a ring graph \cite{Strang1999}.}. If a graph signal $\bm{f} \in \mathbb{R}^{N}$ on vertices of $\mathcal{G}_0$ is downsampled by (GD2), the downsampled signal $\bm{f}_d \in \mathbb{R}^{N/M}$ is equivalent to the signal downsampled using (D1), (D2), or (D3). Furthermore, (GD1) and (GD2) are equivalent if every $M$th sample on the ring graph is retained by (GD1).
\end{corollary}

Similarly, (U2), (GU1), and (GU2) have the following relationship:

\begin{corollary}
Assume $\mathcal{G}_0$ and $\mathcal{G}_1$ are the same as in Corollary \ref{col:downsampling}. If a graph signal $\bm{f} \in \mathbb{R}^{N/M}$ on the vertices of $\mathcal{G}_1$ is upsampled by (GU2) to $\bm{f}_u \in \mathbb{R}^N$, (U1), (U2), (U3), and (GU2) are equivalent. Furthermore, (GU1) and (GU2) are equivalent.
\end{corollary}

In contrast to a ring graph, a path graph does not have such a simple relationship to time domain sampling. Since the DFT is defined by sampling uniform-interval points of the unit circle in the complex plane (i.e., $\omega \in [-\pi, \pi]$), the two bases composed of a conjugate pair cancel each other after spectral domain sampling. In contrast, the DCT basis vectors (eigenvectors of the path graph) are defined on the half-circle $\omega \in [0, \pi]$ to realize a set of real basis functions \cite{Strang1999, Pusche2008}. Hence, the two basis vectors do not cancel each other.

\subsection{Relationship with Sampling Theory for Graph Signals}\label{subsec:samplingtheorem}
Sampling theory for graph signals, i.e., recovering a bandlimited graph signal from a subset of its vertices, has been extensively studied, and many applications have been identified \cite{Pesens2008, Anis2014, Chen2015, Wang2015, Anis2016, Tsitsv2016, Gadde2014, Sakiya2016, Sakiya2017}. In brief, with conventional approaches, the conditions are for recovering the original signal from a subset of samples \textit{after vertex domain sampling}.

As described above, signals sampled using the downsampling method (GD1) do not inherit the desired properties in the graph spectral domain, so, with conventional approaches, asymmetric structures are needed. Although (vertex domain) downsampling is simple, for upsampling, the interpolation or alternating projection algorithms should be carefully designed to recover missing signal values. In contrast, with (GD2), the original signal can be easily reconstructed from its downsampled version in the spectral domain:

\begin{corollary}\label{col3}
If the original graph signal is bandlimited by $\la_{0, N/M}$, i.e., $\widetilde{f}[k] = 0$ for $k > N/M$, it can be perfectly recovered from its downsampled version by using (GD2) if the ideal low-pass filter \eqref{eqn:antialiasingfilter} is applied after using (GU2).
\end{corollary}
This sampling needs only simple ideal filters and thus can be regarded as a counterpart to classical frequency sampling in discrete signal processing.

If the downsampling method (GD3) is used, more careful consideration must be given regarding interpolation of the original spectrum due to differences between the maximum eigenvalues of the original and reduced-size graph Laplacians. This problem remains for future study.

\subsection{Limitations}
\label{subsec:limitation}
The graph signal sampling methods in the spectral domain defined above result in inheritance of the properties of sampling in the frequency domain for classical signal processing. However, in general, they do not preserve the signal values in the vertex domain. It can therefore be said that definitions (GD2), (GD3), (GU2), and (GU3) only partially satisfy the requirements, (P2) and (P3) described in Section \ref{subsec:desirableprop}, for down- and upsampling of signals (see also Table \ref{tb:property}). The vertex domain signals after the proposed downsampling are described in Section \ref{subsec:fracsamp}.

As is well known, the eigenvalue distribution of graph Laplacians greatly depends on the graph, and the eigenvalues are sometimes repeated. This gives freedom to choose the ``best" eigenvectors for the main and aliasing components. Deri and Moura \cite{Deri2017} studied this problem and it could be solved by utilizing their work. Additionally, if neighboring eigenvalues are apart from each other, it is difficult to interpolate the continuous spectrum for (GD3) and (GU3). These remain as challenging problems. The following section describes experimental testing of such effects.

\section{Experimental Results}\label{sec:examples}
As mentioned above, sampling in classical signal processing introduces the same frequency characteristics regardless of whether (D1), (D2), or (D3) or whether (U1), (U2), or (U3) are used. However, (GD1), (GD2), and (GD3) and (GU1), (GU2), and (GU3) have large differences, mainly due to the distribution of the eigenvalues of the graph Laplacians. Examination of toy examples for different graphs can clarify the differences.

For (GD3) and (GU3), an appropriate interpolation method of the discrete spectrum should be used to obtain a spectrum in the continuous domain. However, as previously mentioned, no assumptions concerning the corresponding underlying manifold of $\mathcal{G}$ were made in this study. Therefore, simple linear interpolation for (GD3) and (GU3) is used hereafter. An accurately designed interpolator would improve the sampling performance, and this is left for future work.

\subsection{Sampling on Path Graph}
Intuitively, the size of a path graph can be reduced by dropping every other vertex. Also intuitively, upsampling can be performed by placing a new vertex between every pair of neighboring vertices. Both downsampling and upsampling for the path graph are therefore described here.

First, a downsampling example for a path graph signal is described. The spectral response with downsampling by two for (GD1)--(GD3) with $N = 100$ is shown in Fig. \ref{fig:examples_downsampling}(a). The bandwidth of the graph signal was set sufficiently narrow to prevent aliasing. The spectral response with (GD1) has an undershoot at high frequency and gradually decays. The spectral responses with (GD2) and (GD3) do not exhibit such effects and instead exhibit different characteristics. That is, with (GD2), the bandwidth is relatively broad because of the eigenvalue distribution of the path graph. The eigenvalues are densely distributed in the low- and high- frequency regions, whereas they are relatively sparse in the mid-frequency region.

For upsampling, a graph signal with a narrow bandwidth is also considered. The original graph has $N=50$ vertices, and the signal is upsampled by two. After upsampling, there should be one main and one imaging component. As shown in Fig. \ref{fig:examples_downsampling}(b), despite the bandlimited input, the spectrum upsampled using (GU1) has a nonzero response at all eigenvalues. As for downsampling, the use of (GU2) and (GU3) led to the expected spectral response, but the bandwidths depend on the distribution of the eigenvalues, similar to the downsampling case.

\subsection{Sampling on Grid Graph}
Next, a downsampling example for a 2-D grid graph is described. The original graph has $N = 256$ vertices evenly distributed in 2-D space $[0, 1) \times [0, 1)$. The reduced-size graph has $N = 64$ vertices, i.e., $M = 4$. As in the case of the path graph, it is easy to determine the remaining vertices after downsampling. One vertex in every four is selected. The spectra are shown in Fig. \ref{fig:examples_downsampling}(c). As expected, they are stretched by four on the spectrum index domain (for downsampling using (GD2)) or the width of the spectrum (for downsampling using (GD3)). In contrast, the downsampled spectrum no longer has the expected characteristics when (GD1) is used.

\subsection{Sampling on Random Regular Graph}
Consider a signal on a random regular graph with $N = 100$ in which every vertex is randomly connected to ten other vertices. The smallest eigenvalue of the graph Laplacian is further apart from the other. A reduced-size graph with $N = 50$ is constructed on the basis of the method presented by Shuman et al. \cite{Shuman2016}; that is, it is no longer a regular graph due to the requirement of the one-to-one vertex mapping of (GD1)\footnote{In contrast, (GD2) and (GD3) do not require such one-to-one mapping. Examples are described in Section \ref{subsec:fracsamp}.}. The spectra of the downsampled signals are shown in Fig. \ref{fig:examples_downsampling}(d). Although the eigenvalues of the graph Laplacian are not distributed uniformly, the spectral characteristics due to downsampling can still be considered reasonable.

\begin{figure}[tp]%
\centering
\subfigure[][Example of (GD2).]{\includegraphics[width=\linewidth]{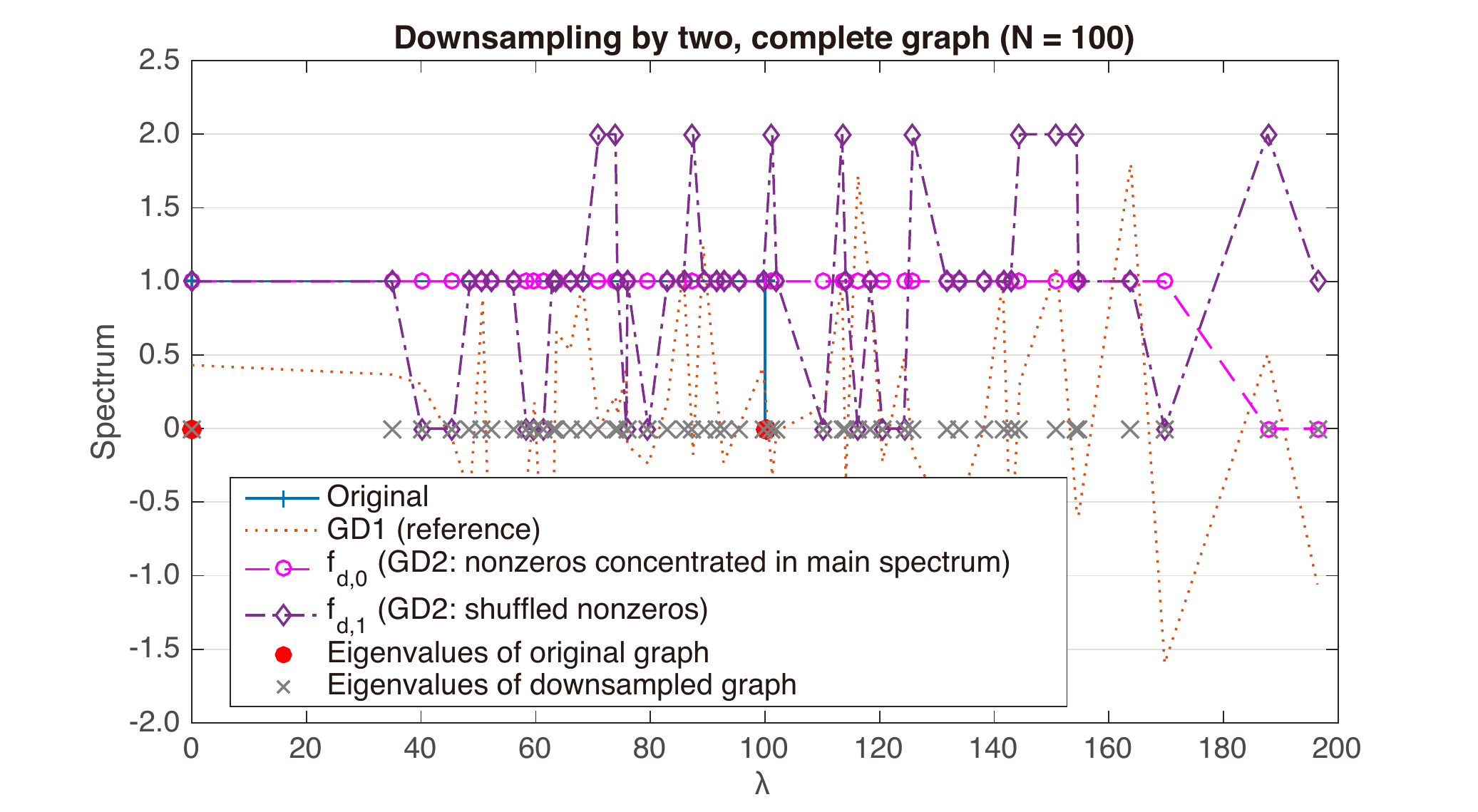}}\\
\subfigure[][Example of (GD3).]{\includegraphics[width=\linewidth]{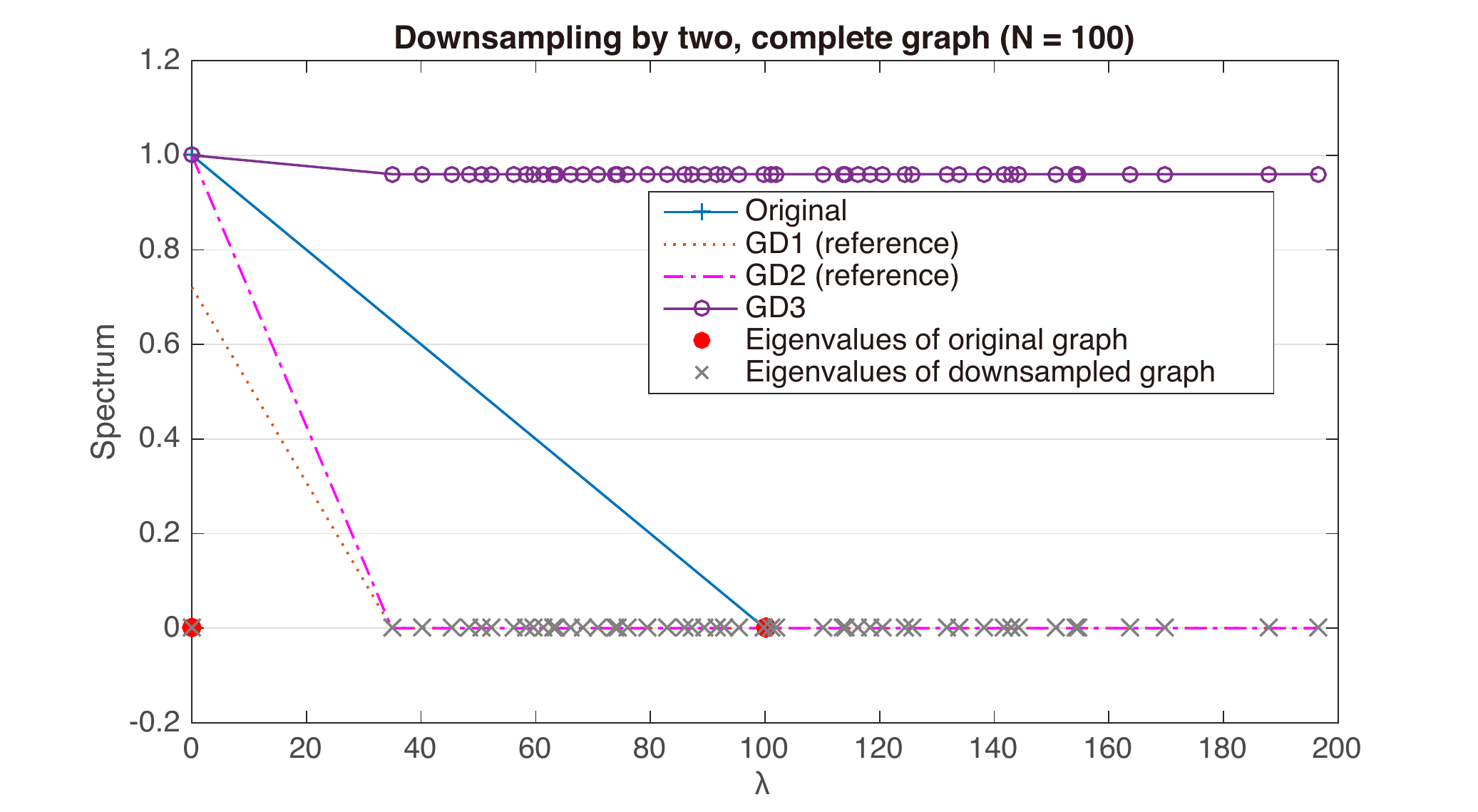}}
\caption{Effects of repeated eigenvalues. Graph used is complete graph with $|\mathcal{V}_0| = 100$; reduced-size graph is made using Kron reduction \cite{Shuman2016, Dorfle2013} ($|\mathcal{V}_1| = 52$). (a) Example spectral response with (GD2). Original signal has $50$ nonzero spectrum. $\bm{f}_{d,0}$ is the downsampled spectrum of which the nonzero spectrum is concentrated in the first $50$ coefficients, whereas those in $\bm{f}_{d,1}$ are randomly permuted. (b) Example spectral response with (GD3); $\widetilde{f}[0] = 1$ and remaining coefficients are zero.}
\label{fig:repeated_eigenvalue}
\end{figure}

\subsection{Aliasing Effects}\label{subsec:aliasing}
The aliasing effects for non-bandlimited graph signals with (GD2), (GD2'), (GD3), and (GD3') are compared in Fig. \ref{fig:path_aliasing} for a path graph with $N=100$. The signal is downsampled by two as in the previous example. However, the graph signal is \textit{neither} bandlimited from the index nor the spectral perspectives. Though (GD2) has a close relationship with (D1), (D2), and (D3), the aliasing effect is significant even when the spectra have little overlap; that is, (GD2) affects the spectrum at low frequencies. A similar effect occurs for (GD3). In contrast, if the modified downsampling methods (GD2') and (GD3') are used, their effects on the spectrum are only slight, as expected. Therefore, to avoid an unexpected large side effect, (GD2') and (GD3') should be used.

\begin{figure}[tp]%
\centering
\subfigure[][]{\includegraphics[width=.8\linewidth]{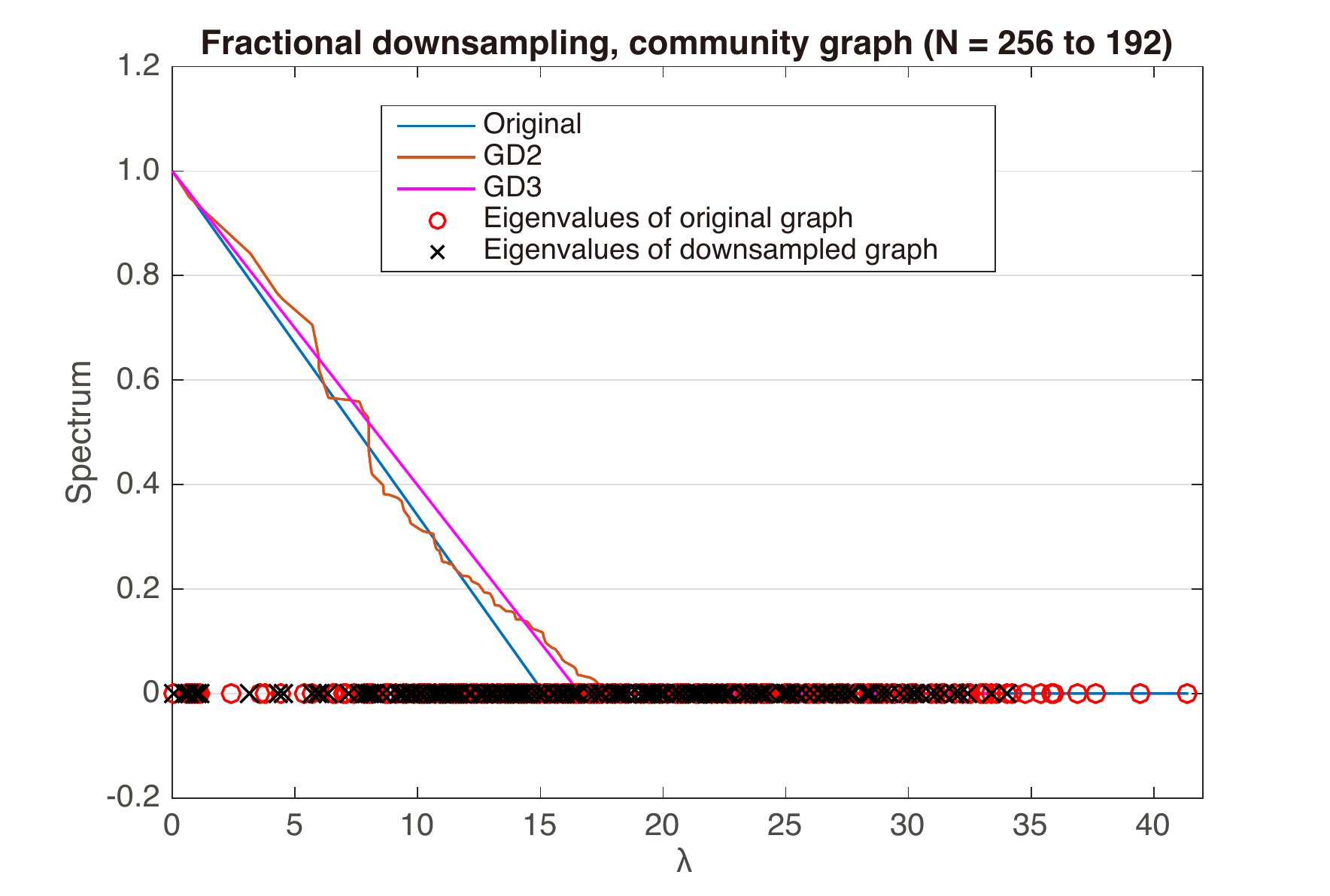}}\\
\subfigure[][]{\includegraphics[width=\linewidth]{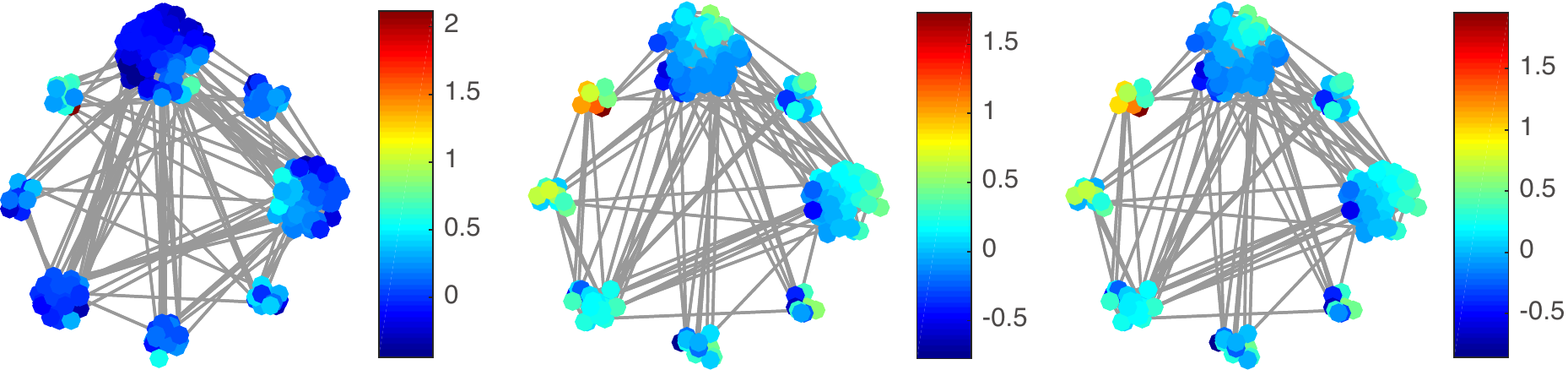}}
\caption{Fractional downsampling of signal on \textit{Community Graph} with eight communities ($|\mathcal{V}_0| = 256$ to $|\mathcal{V}_1| = 192$). Downsampling ratio is $4/3$. Reduced-size graph also has eight communities, but original and reduced-size graphs do not have a one-to-one relationship. (a) Spectrum of original and downsampled signals. (b) Signals in vertex domain. From left to right: original signal, signal downsampled using (GD2), and signal downsampled using (GD3).}
\label{fig:comm_frac_samp}
\end{figure}

\begin{figure}[tp]%
\centering
\subfigure[][]{\includegraphics[width=.8\linewidth]{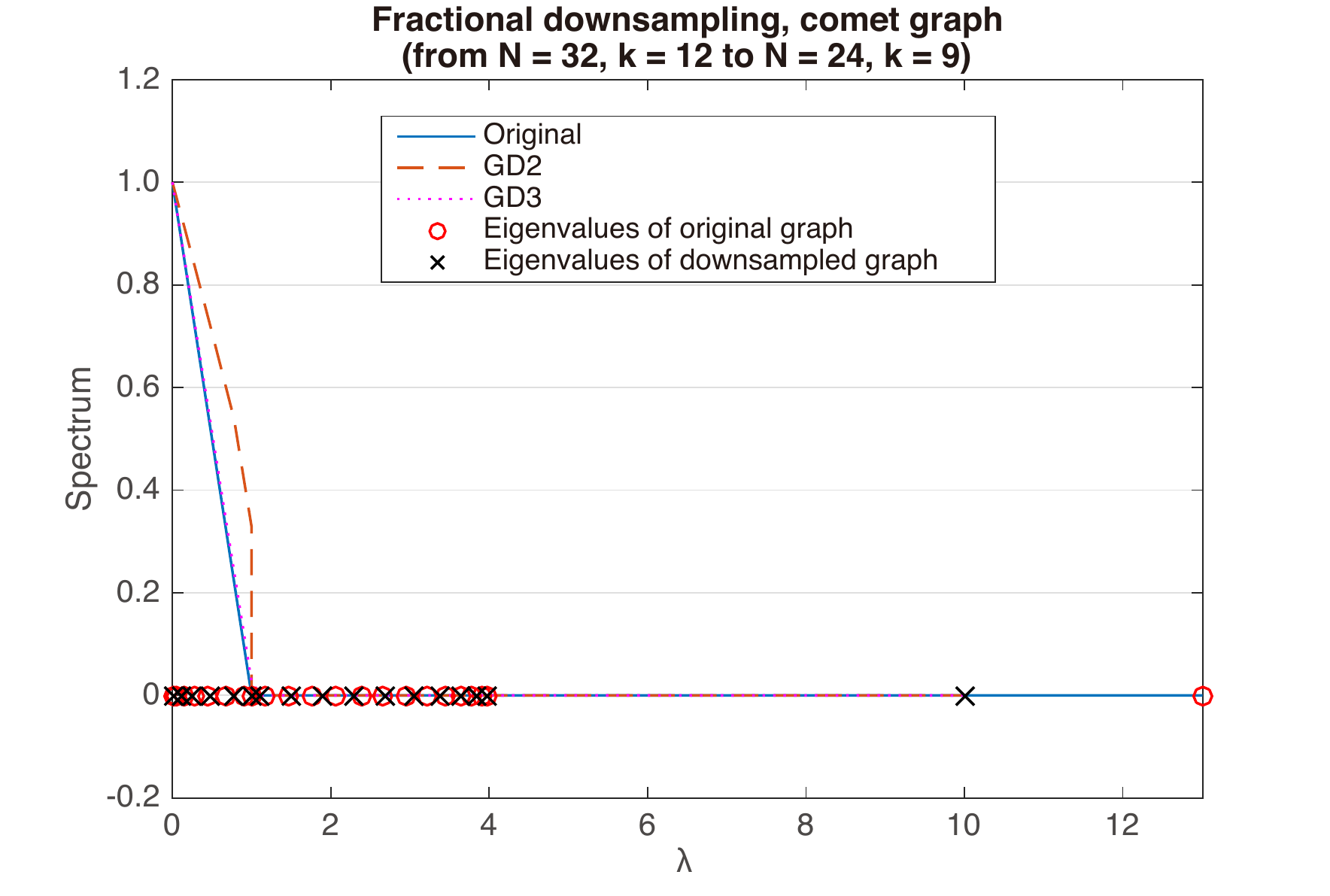}}\\
\subfigure[][]{\includegraphics[width=.8\linewidth]{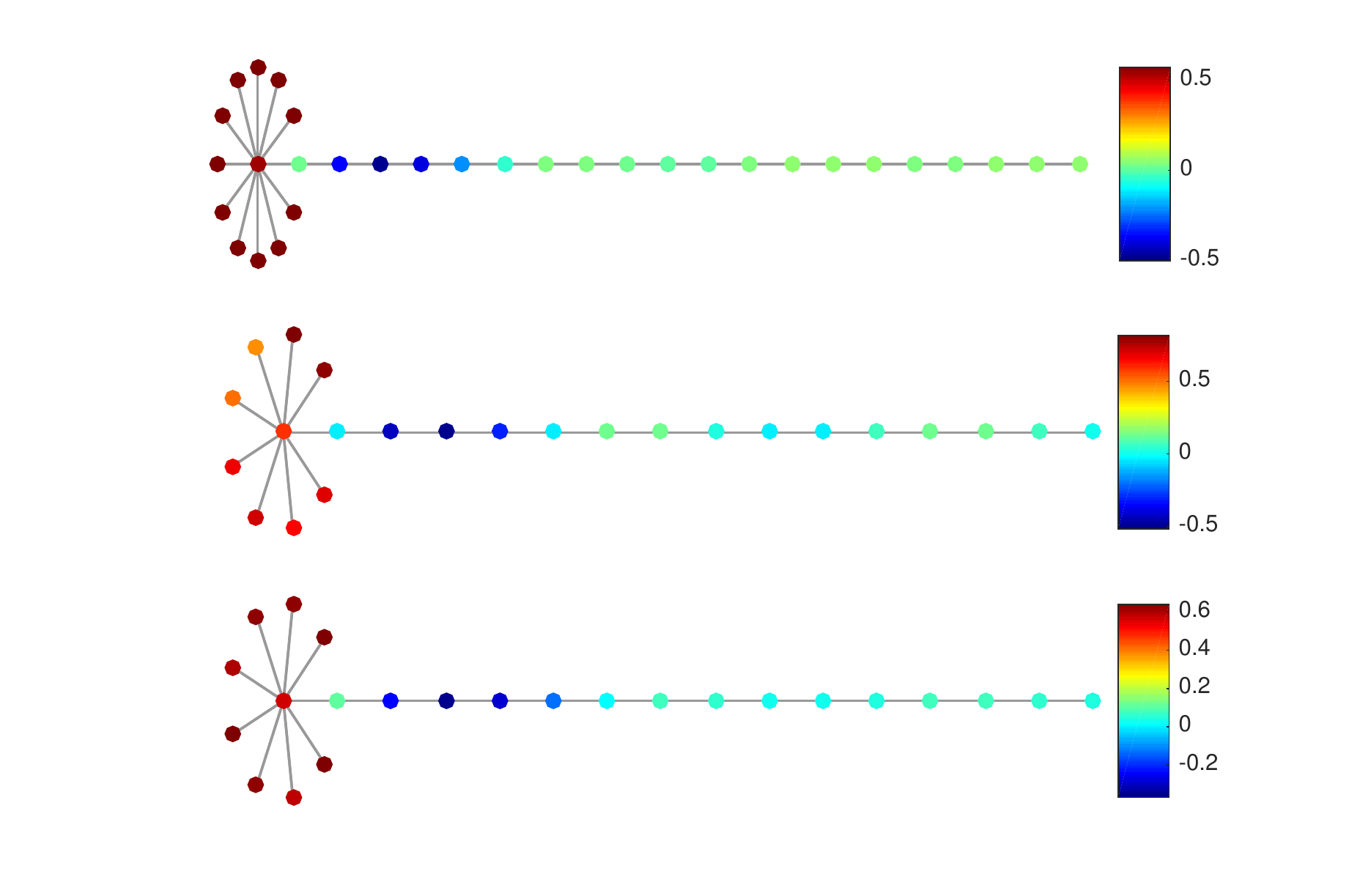}}
\caption{
Fractional downsampling of signal on \textit{Comet Graph} ($|\mathcal{V}_0| = 32$ to $|\mathcal{V}_1| = 24$). Original graph has $32$ vertices and degree of center vertex is $12$; reduced-size graph has $24$ vertices and degree of center vertex is $9$ (i.e., downsampling ratio is $4/3$). (a) Spectra of original and downsampled signals. (b) Signals in vertex domain (from top to bottom): original signal, signal downsampled using (GD2), and signal downsampled using (GD3).}
\label{fig:comet_frac_samp}
\end{figure}

\begin{figure*}[tp]%
\centering
\includegraphics[width=\linewidth]{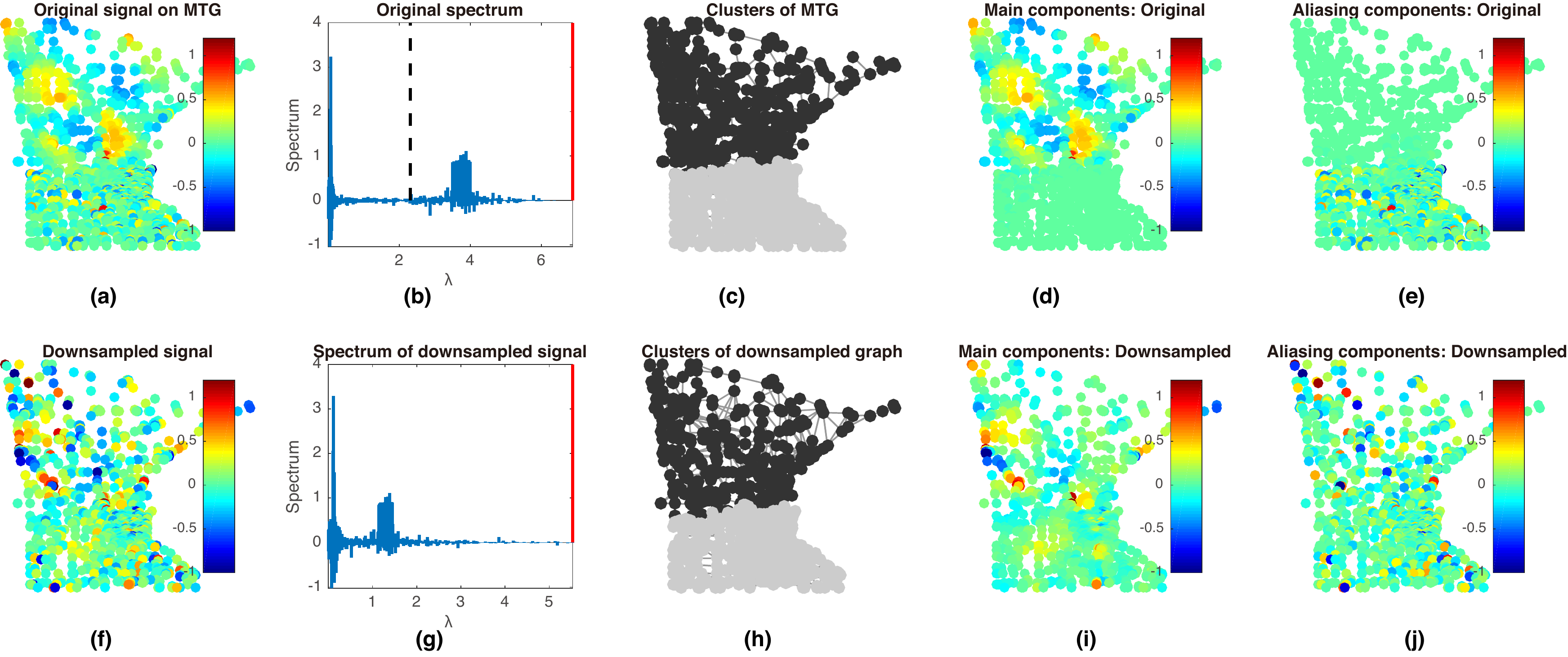}
\caption{Spectral domain downsampling for graph signal on Minnesota Traffic Graph ($|\mathcal{V}_0| = 2642$ to $|\mathcal{V}_1| = 1323$). Downsampling method is (GD2). Top ((a)--(e)): original signal. Bottom ((f)--(i)): downsampled signal. First columns ((a) and (f)): original and downsampled signals in vertex domain. Second columns ((b) and (g)): spectra of original and downsampled signals; red lines indicate maximum eigenvalues and black dashed line represents $\la_{0, 1323}$, i.e., folding point of spectrum. Third columns ((c) and (h)): vertex clusters. Fourth columns ((d) and (i)): main components of original and downsampled signals. Fifth columns ((e) and (j)): aliasing components of original and downsampled signals.}
\label{fig:minnesota_sampling}
\end{figure*}

\subsection{Effects of Repeated Eigenvalues}\label{subsec:repeatedev}
Consider the extreme situation in which there are repeated eigenvalues, as described in Section \ref{subsec:limitation}, due to downsampling of signals on a complete graph $\mathcal{K}_N$. The smallest eigenvalue of the graph Laplacian is known to be $0$ and all of the remaining $(N-1)$ eigenvalues are known to be the same and equal to $N$. The eigenvalue distribution is also extreme.

In this case, we encounter the following problems;
\begin{itemize}
 \item For (GD2): There is freedom to select the eigenvectors corresponding to the main and folding spectra.
 \item For (GD3): The continuous spectrum has to be interpolated from only two distinct values, $\widetilde{f}[0]$ and $\widetilde{f}[i]$, where $i = 1, \ldots, N-1$.
\end{itemize}

In this example $N=100$, and the reduced-size graph $\mathcal{G}_1$ is calculated from $\mathcal{K}_{100}$ using Kron reduction \cite{Shuman2016, Dorfle2013}. That is, although $\mathcal{G}_1$ is no longer a complete graph, all vertices in $\mathcal{G}_1$ correspond to vertices in $\mathcal{K}_{100}$.

\begin{figure}[t]%
\centering
\subfigure{\includegraphics[width=.8\linewidth]{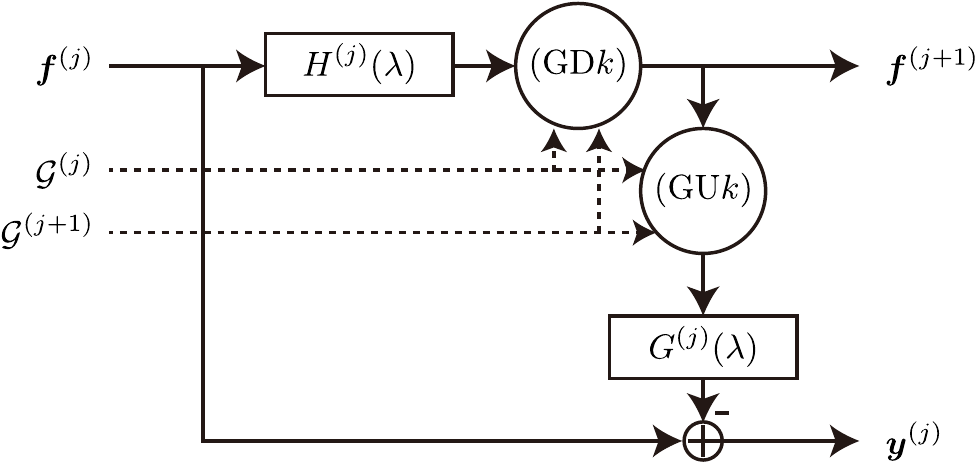}}\\
\subfigure{\includegraphics[width=.8\linewidth]{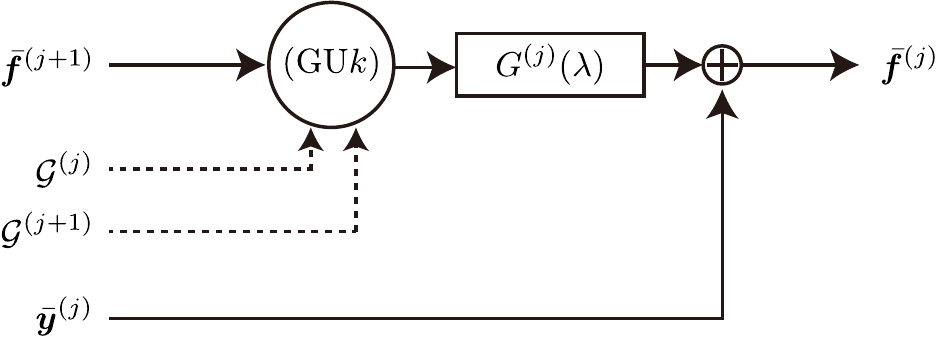}}
\caption{Graph Laplacian pyramid at $j$th level: $\bm{f}^{(j)}$ is $j$th-level input, and $\bm{y}^{(j)}$ is prediction error at $j$th level. For multiscale decomposition, $\bm{f}^{(j+1)}$ is further decomposed using $(j+1)$th level pyramid. (GD$k$) and (GU$k$) ($k \in \{1, 2, 3\}$) are downsampling and upsampling operators of graph signals introduced in this paper, respectively. $H^{(j)}(\la)$ and $G^{(j)}(\la)$ are arbitrary low-pass graph filters for approximation and prediction, respectively. $\mathcal{G}^{(j)}$ is graph for $j$th-level decomposition. Top: analysis transform; bottom: synthesis transform.}
\label{fig:graphLP}
\end{figure}

In the example for (GD2), $50$ samples in $\widetilde{\bm{f}}$ are set to be nonzero ($=1$). Aliasing is avoided by selecting the order of the eigenvectors corresponding to $\la_{\max} = N$ so that all nonzero coefficients appear in the main spectrum, i.e.,
\begin{equation}
\label{ }
\widetilde{\bm{f}}_{0} = [\underbrace{1, \ldots, 1}_{50 \text{ samples}}, \underbrace{0, \ldots, 0}_{50 \text{ samples}}]^\top.
\end{equation}
Conversely, if the nonzero values are assigned to the folding spectrum given the freedom for selecting eigenvectors, aliasing should occur.

A new signal $\widetilde{\bm{f}}_{1} = \bar{\mathbf{U}}^\top_{\mathcal{K}}\mathbf{U}_{\mathcal{K}}\widetilde{\bm{f}}_{0}$ is also tested where $\mathbf{U}_{\mathcal{K}}$ is the eigenvector matrix of $\mathcal{K}_{100}$ and $\bar{\mathbf{U}}_{\mathcal{K}}$ is the reordered version of $\mathbf{U}_{\mathcal{K}}$; the eigenvectors corresponding to $\la_{\max}$ are randomly permuted.

The signals sampled using (GD2) are shown in Fig. \ref{fig:repeated_eigenvalue}(a) along with that using (GD1) for reference\footnote{In this and subsequent examples, (GD3) produces the same spectrum when the signal is bandlimited using a low-pass filter.}. The signals $\bm{f}_{d, 0}$ and $\bm{f}_{d, 1}$ correspond to the downsampled versions of $\bm{f}_{0}$ and $\bm{f}_{1}$, respectively. It is clear that $\bm{f}_{d, 0}$ has no aliasing whereas $\bm{f}_{d, 1}$ randomly yields aliasing.

In the example for (GD3), the bandlimited graph signal considered is $\widetilde{\bm{f}}_2 = [1, 0, \ldots, 0]^\top$. That means $\bm{f}_2$ is a constant signal and (GD1) yields a bandlimited spectrum even after downsampling. The interpolation method used to obtain $\widetilde{f}_{\text{int}}(\la)$ in \eqref{eqn:fint} for (GD3) is critical since the continuous spectrum has to be interpolated from only two values. When linear interpolation is used, $\widetilde{f}_{\text{int}}(\la)$ is nonzero except for $\widetilde{f}_{\text{int}}(\la_{0, \max})$. The example is shown in Fig. \ref{fig:repeated_eigenvalue}(b), where (GD3') is used; severe aliasing is still evident.

The examples above show that the proposed sampling method causes aliasing if there are repeated eigenvalues and/or if the eigenvalue distribution is heavily biased. Studying such extreme cases remains an open problem. In the following section, a few applications are introduced, and spectral domain sampling is shown to work well in not-so-extreme situations.

\begin{figure}[t]%
\centering
\subfigure{\includegraphics[width=.55\linewidth]{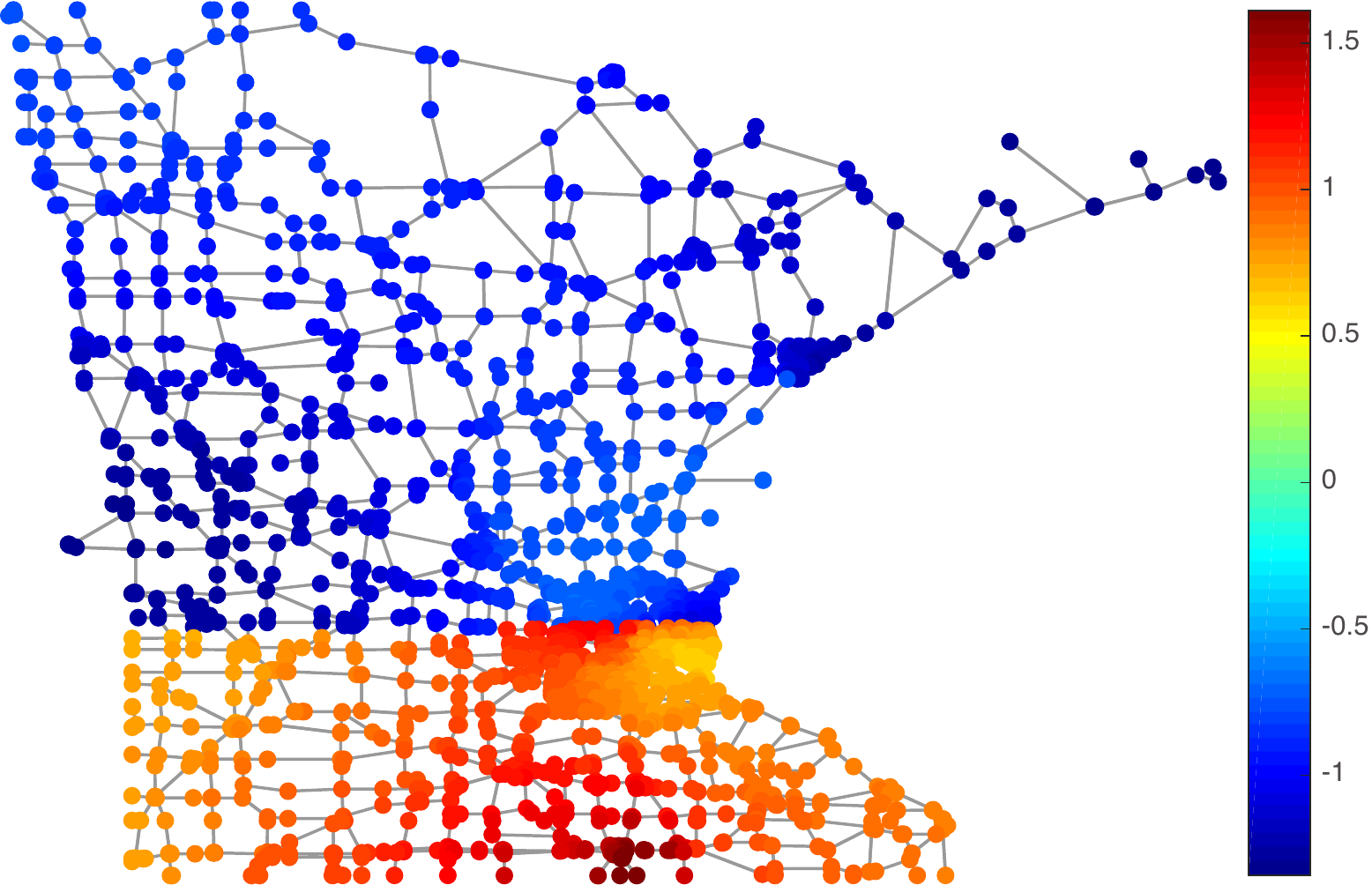}}\ 
\subfigure{\includegraphics[width=.4\linewidth]{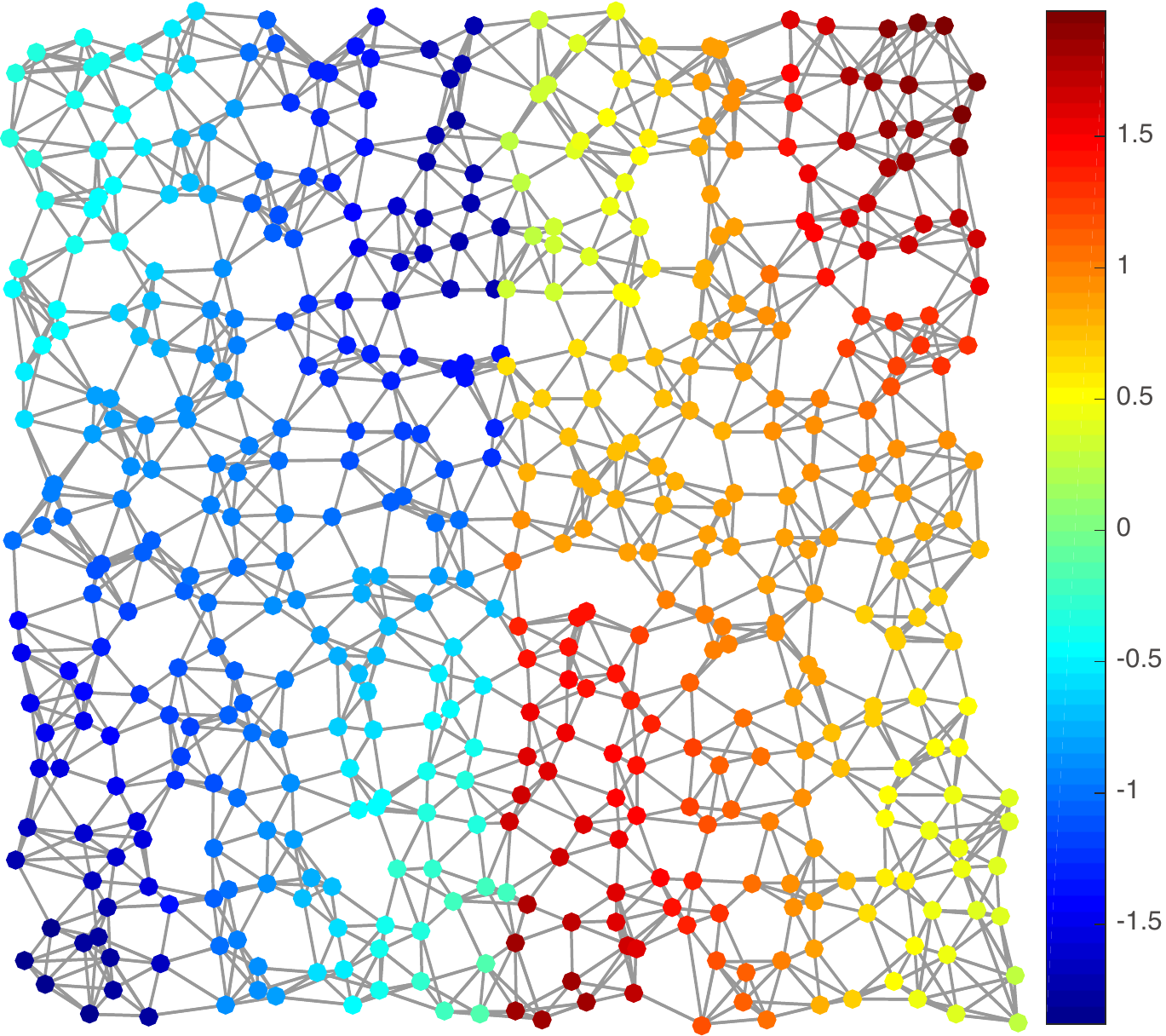}}
\caption{Input signals on graphs. (Left) Graph signal on Minnesota Traffic Graph. (Right) Graph signal on random sensor graph.}
\label{fig:inputsignals}
\end{figure}

\begin{figure*}[tp]%
\centering
\subfigure[][Graph Laplacian pyramid with (GD1) and (GU1)]{\includegraphics[width=.7\linewidth]{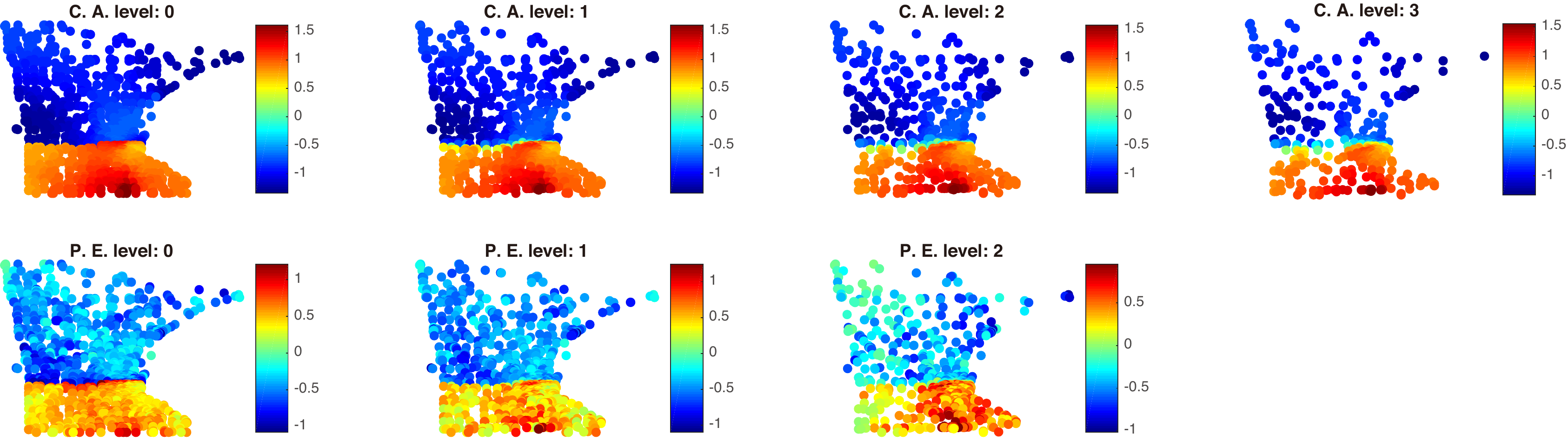}}\\
\subfigure[][Graph Laplacian pyramid with (GD2) and (GU2)]{\includegraphics[width=.7\linewidth]{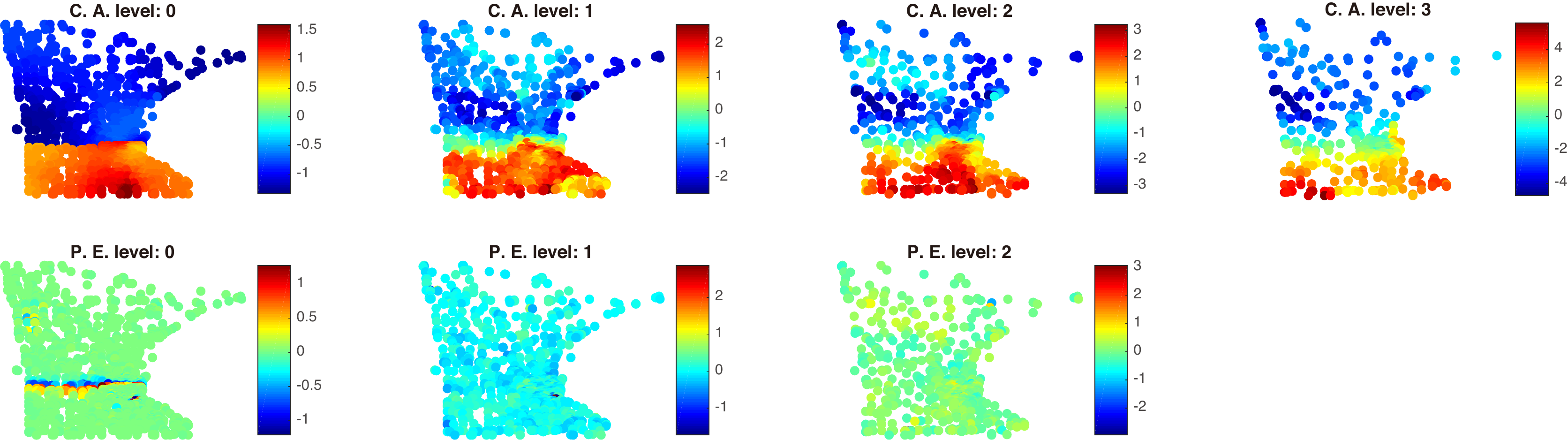}}\\
\subfigure[][Graph Laplacian pyramid with (GD3) and (GU3)]{\includegraphics[width=.7\linewidth]{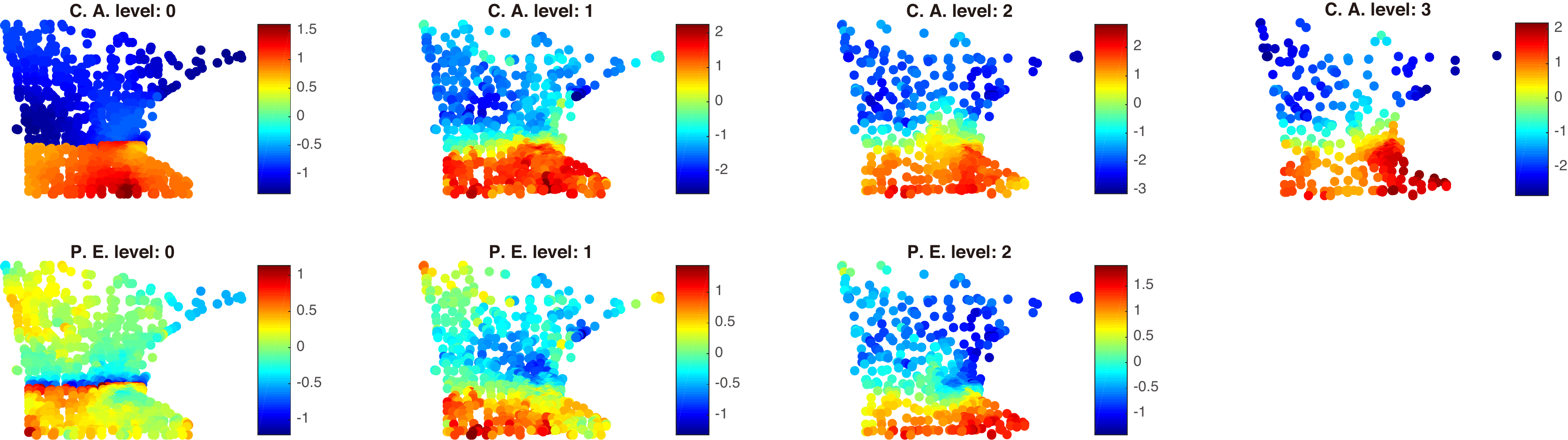}}\\
\caption{Decomposed signals using graph Laplacian pyramid for various levels. Decomposition level is three. Top row: coarse approximations ($\bm{f}^{(j)}$ in Fig. \ref{fig:graphLP}), bottom row: prediction errors ($\bm{y}^{(j)}$ in Fig. \ref{fig:graphLP}).}
\label{fig:GLP_decomposition}
\end{figure*}

\section{Potential Applications}\label{sec:applications}
There are various potential applications of sampling on the graph spectral domain. As with the proposed method, downsampling and upsampling operators (GD$k$') and (GU$k$') ($k \in \{2, 3\}$) are used here instead of (GD$k$) and (GU$k$) since they show slightly better performance.

\subsection{Fractional Sampling}\label{subsec:fracsamp}
Fractional sampling is useful in classical signal processing and would be useful in graph signal processing for reducing (increasing) the number of samples while avoiding aliasing (imaging). If downsampling (upsampling) could be performed while obtaining spectral domain effects similar to those for the frequency domain, the proposed spectral domain sampling methods would be suitable for fractional sampling.

For simplicity, fractional downsampling is considered here. In the classical case, the input signal is first upsampled by $L$ and then downsampled by $M$ to obtain a downsampled signal with rate $M/L$. However, in the graph setting, this approach would be problematic if used straightforwardly because three graph Laplacians would be involved: the original, the upsampled, and the downsampled ones. Designing the upsampled (oversampled) one is especially difficult since an increased-size graph Laplacian has to be estimated.

A different approach can be used instead: stretch the spectrum by $M/L$. This approach is slightly different from that used for (GD2) and (GD3) as described in Section \ref{sec:spectral_samp}, but it is a natural extension.

Interestingly, in contrast to most of the vertex domain approaches, spectral domain sampling does not necessarily have one-to-one vertex mapping between the original and the reduced-size graphs. This is because the proposed sampling approach \textit{translates} the original spectral information into a different graph rather than copying the signal value itself.

\subsubsection{Community Graph}
An example of fractional sampling of a signal on a community graph is shown in Fig. \ref{fig:comm_frac_samp}. The original graph has eight communities and $|\mathcal{V}_0| = 256$. The signal is downsampled to $|\mathcal{V}_1| = 192$, i.e., the downsampling ratio is $4/3$. The reduced-size graph also has eight communities, but the original and reduced-size graphs do not have a one-to-one relationship. Though the signal values are not equal to the original signal, they are fairly close.

\subsubsection{Comet Graph}
Fractional sampling of a signal on a comet graph can also be performed. For the original comet graph, $|\mathcal{V}_0| = 32$ is set, and the degree of the center vertex is set to $12$, whereas for the reduced-size graph, $|\mathcal{V}_1|=24$, and the degree of the center vertex is set to $9$. The resulting downsampling ratio is $4/3$. The signal is set to be bandlimited. The spectra of the original and downsampled signals are shown in Fig. \ref{fig:comet_frac_samp}(a). These in the vertex domain are also shown in Fig. \ref{fig:comet_frac_samp}(b). It is clear that the downsampled signals have characteristics similar to those of the original one even in the vertex domain.

\subsubsection{Minnesota Traffic Graph}
The behavior of the main and aliasing components are discussed here for a test signal on the Minnesota Traffic Graph ($|\mathcal{V}_0| = 2642$) shown in Fig. \ref{fig:minnesota_sampling}(a). The signal is downsampled using (GD2) (shown in Fig. \ref{fig:minnesota_sampling}(f)). The test signal is obtained using the same two steps used by Shuman et al. \cite{Shuman2015}: 1) the vertex set is divided into two clusters by spectral clustering; 2) each cluster is forced to have one (main or aliasing) spectral component. Formally, the signal is defined as $\bm{f} = \bm{f}_1/||\bm{f}_1||_\infty + \bm{f}_2/||\bm{f}_2||_\infty$, where
\begin{equation}
\label{ }
f_j[n] := \mathbbm{1}_{\{\text{vertex }n \text{ is in cluster }j\}}\sum_{k=0}^{N-1}u_k[n]\mathbbm{1}_{\{\underline{\tau}_j \leq \la_k \leq \bar{\tau}_j\}}.
\end{equation}
In Fig. \ref{fig:minnesota_sampling}, $\{[\underline{\tau}_1, \bar{\tau}_1], [\underline{\tau}_2, \bar{\tau}_2]\}$ is set to $\{[0.06, 0.08], [3.5, 4.0]\}$ for the dark gray and light gray clusters, respectively.

The reduced-size graph $\mathcal{G}_1$ is calculated using Kron reduction \cite{Shuman2016, Dorfle2013}, where $|\mathcal{V}_1| = 1323$. The original spectrum is folded at $\la_{0, 1323} = 2.31$, as shown by the black dashed line in the original spectrum (Fig. \ref{fig:minnesota_sampling}(b)). This leads to the dark gray cluster corresponding to the main spectrum and the light gray cluster corresponding to the aliasing spectrum, as illustrated in Figs. \ref{fig:minnesota_sampling}(d) and (e).

The vertex and spectral domain signals after using (GD2) are shown in Figs. \ref{fig:minnesota_sampling}(f) and (g), respectively. As previously mentioned, the vertex signal values are not preserved; however, the energy in the main spectrum is still concentrated in the dark gray cluster, which is shown in Fig. \ref{fig:minnesota_sampling}(i). Numerically, $||\bm{f}^{\text{main}}_{d, \text{ Cluster }1}||^2_2 = 35.2$, whereas $||\bm{f}^{\text{main}}_{d, \text{ Cluster }2}||^2_2 = 15.3$, where $f^{\text{main}}_{d, \text{ Cluster }k}[n] = \mathbbm{1}_{\{\text{vertex } n\text{ is in cluster }k\}} \cdot\widetilde{f}_d[n]$. In contrast, the energy in the aliasing spectrum is distributed to both clusters (Fig. \ref{fig:minnesota_sampling}(j)): $||\bm{f}^{\text{alias}}_{d, \text{ Cluster }2}||^2_2 = 30.9$ and $||\bm{f}^{\text{alias}}_{d, \text{ Cluster }2}||^2_2 = 29.3$. This is because the folded aliasing spectrum becomes a lower frequency component that no longer corresponds to the second cluster. Adaptive selection of the folding indices should improve the vertex/spectral domain properties.

\begin{figure}[t]%
\centering
\subfigure[][]{\includegraphics[width=.8\linewidth]{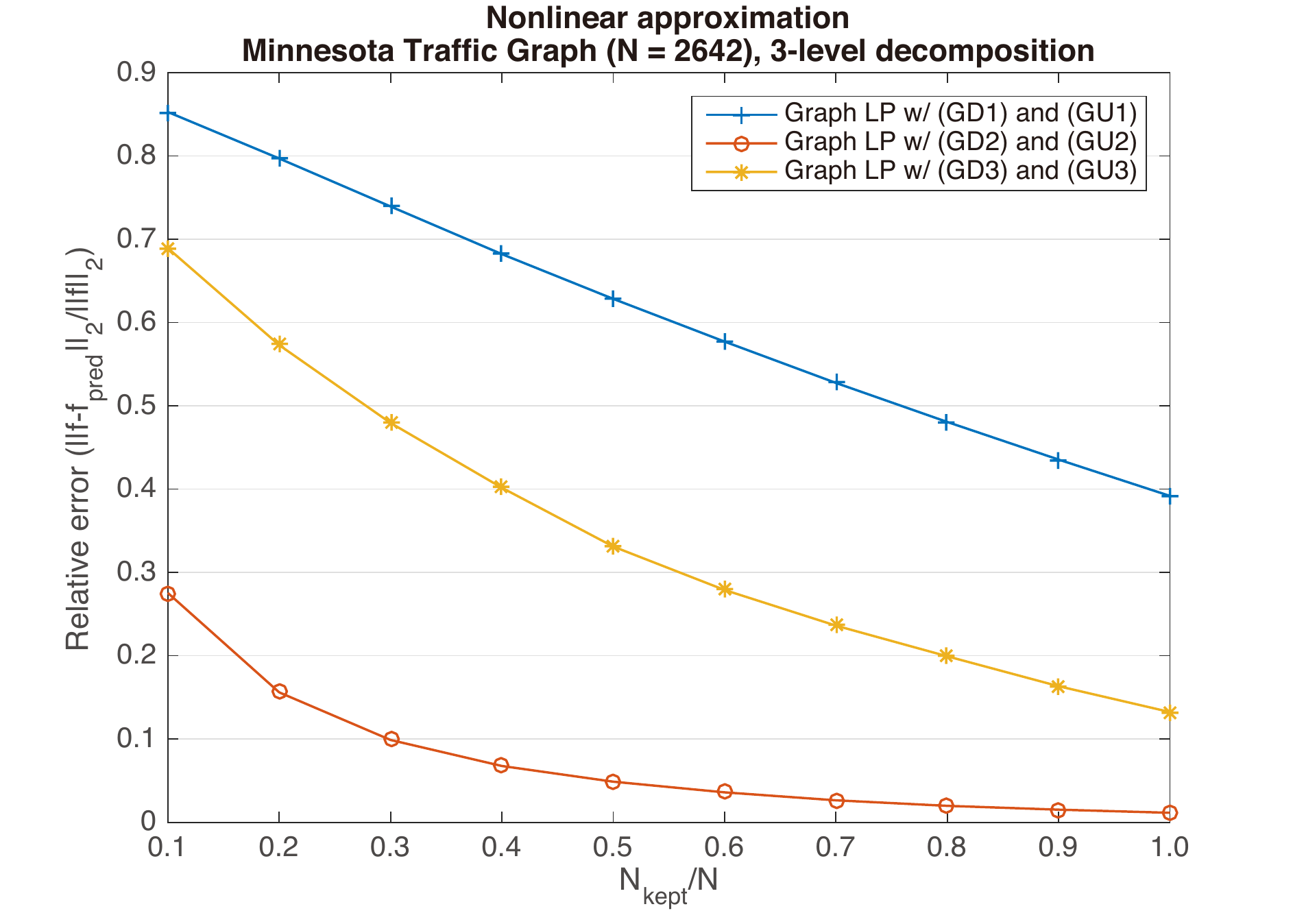}}\\
\subfigure[][]{\includegraphics[width=.8\linewidth]{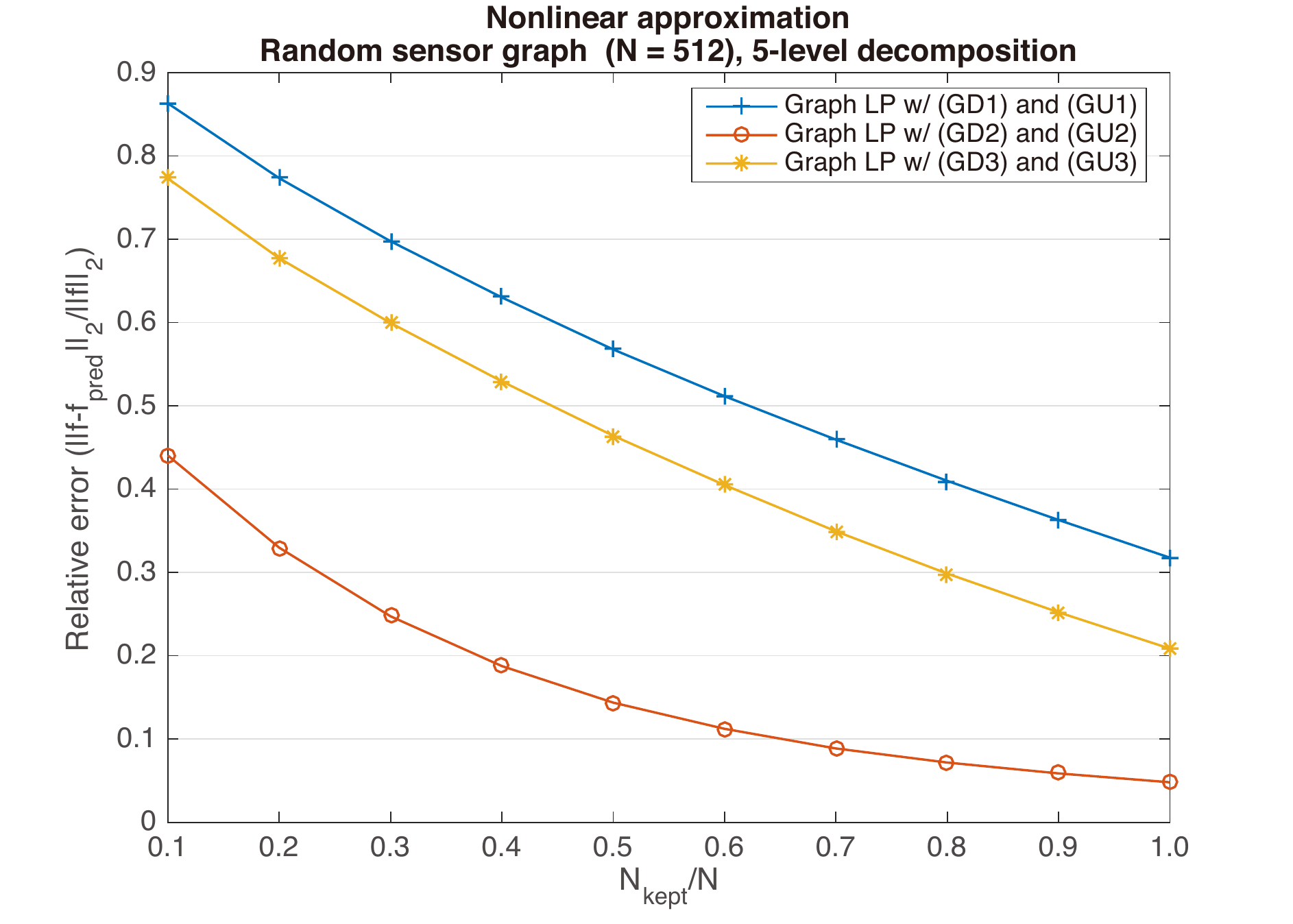}}
\caption{Nonlinear approximation with graph Laplacian pyramid. All coefficients in the coarsest subband are retained, and $N_\text{kept}$ coefficients having the largest magnitude in the high-frequency subband are kept for reconstruction. Remaining coefficients are set to zero. Normalized errors $||\bm{f} - \bm{f}_\text{pred}||_2/||\bm{f}||_2$ in accordance with fraction of retained coefficients $N_\text{kept}/N$ are compared. (a) Minnesota Traffic Graph and (b) random sensor graph.}
\label{fig:NLA_error}
\end{figure}

\begin{figure}[t]%
\centering
\subfigure[][]{\includegraphics[width=.5\linewidth]{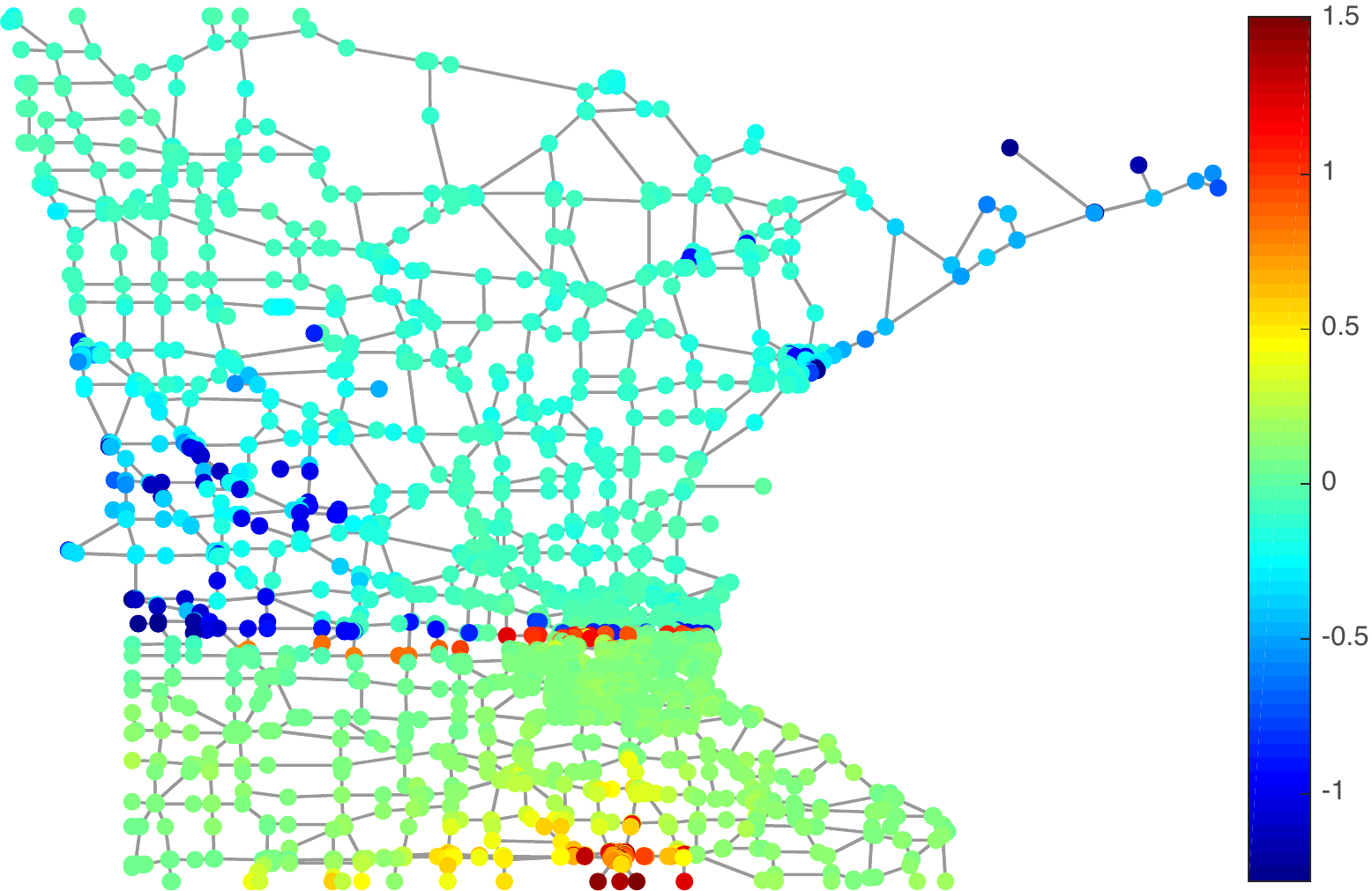}}\\
\subfigure[][]{\includegraphics[width=.5\linewidth]{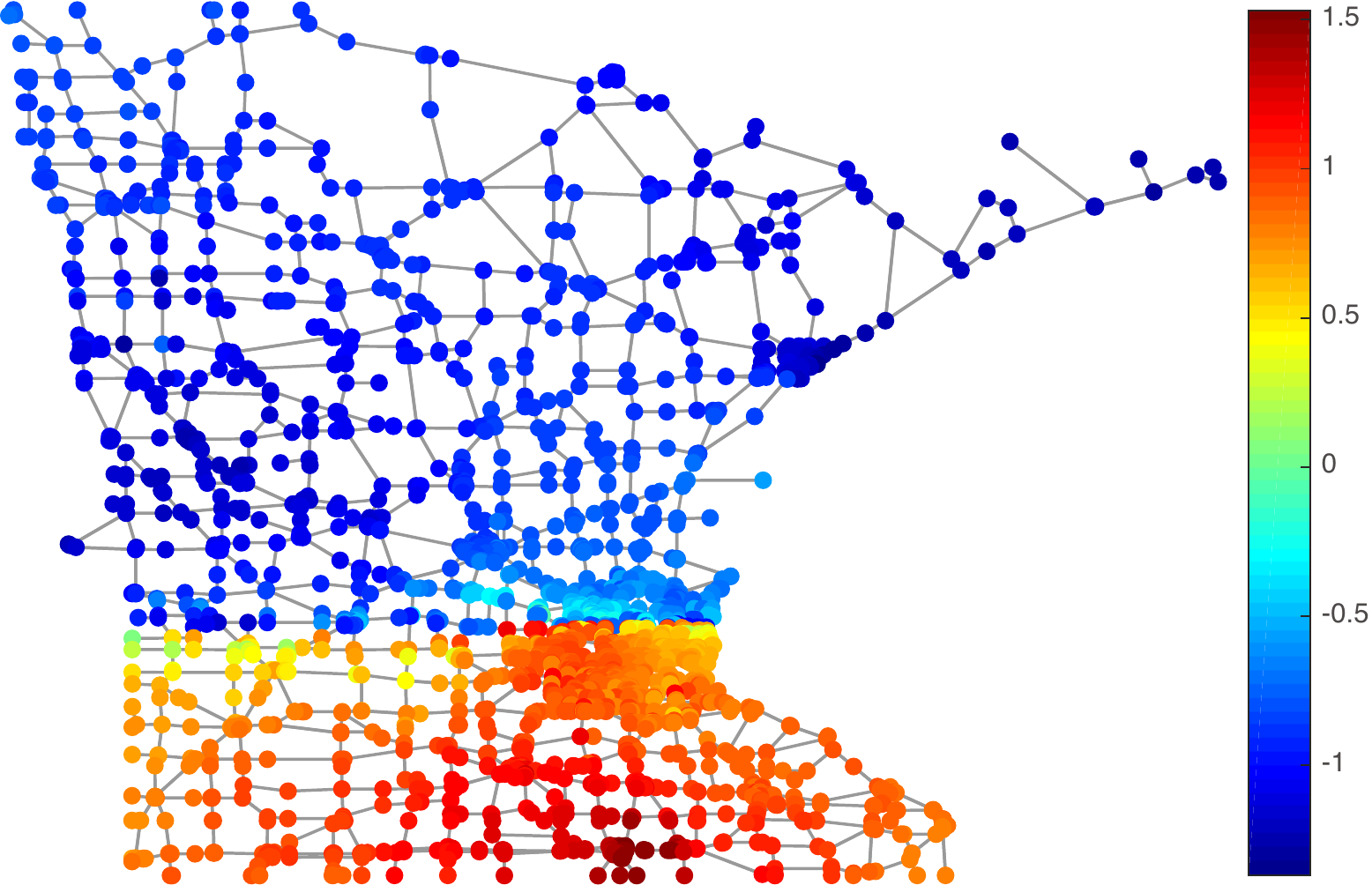}}\\
\subfigure[][]{\includegraphics[width=.5\linewidth]{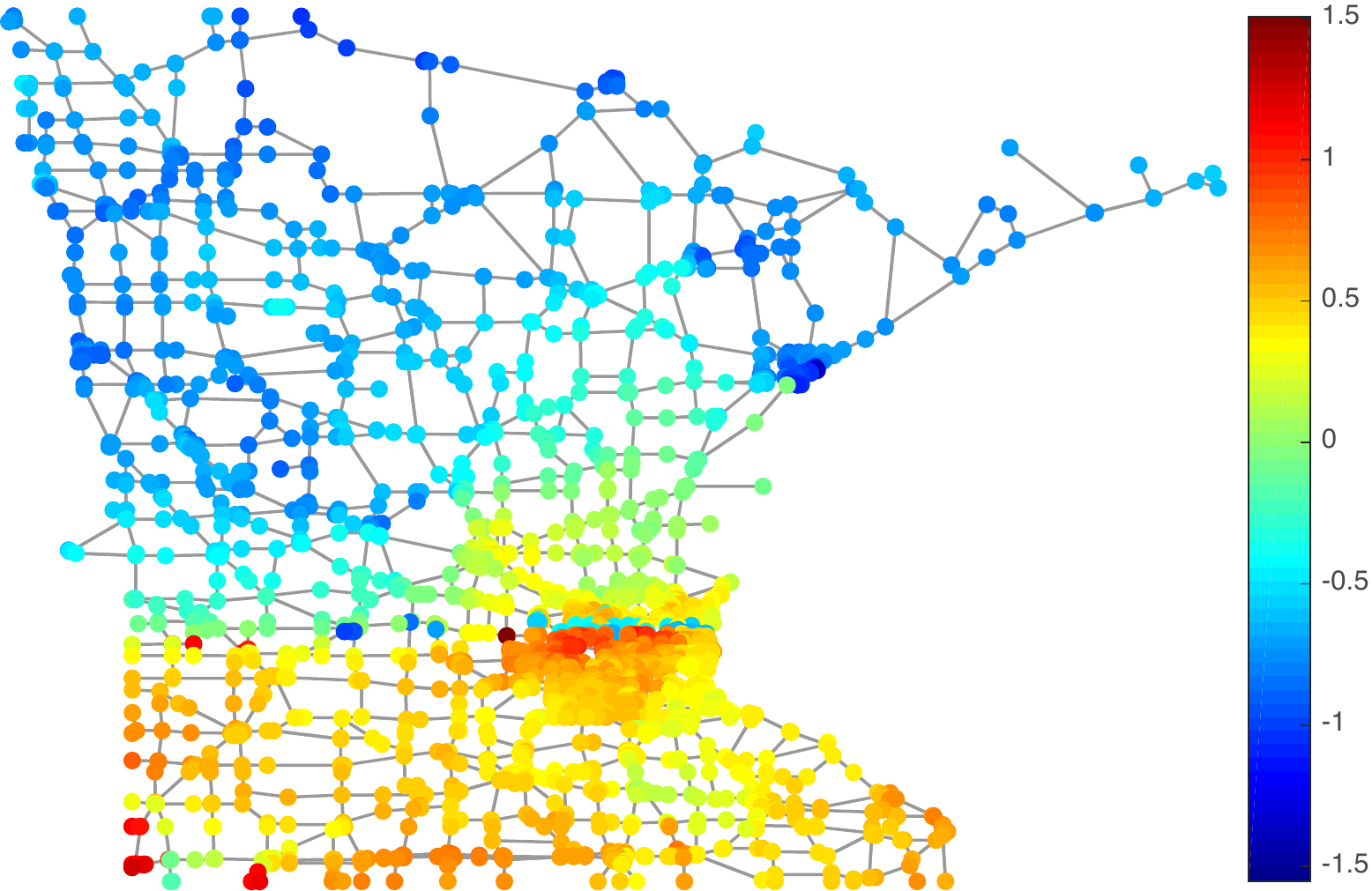}}
\caption{Nonlinear approximation with graph Laplacian pyramid. $N_{\text{kept}} = 0.2N.$ From top to bottom: signal reconstructed using (GD1) and (GU1), signal reconstructed using (GD2) and (GU2), and signal reconstructed using (GD3) and (GU3).}
\label{}
\end{figure}

\subsection{Graph Laplacian Pyramid}
To demonstrate the effectiveness of the spectral domain sampling, the proposed methods were applied to a Laplacian pyramid for graph signals \cite{Shuman2016}. This yielded the coarsest approximation and several subband signals at finer scales.

As shown in Fig. \ref{fig:graphLP}, downsampling and upsampling were performed at each decomposition level, so the sampling methods can be compared. In the study by Shuman et al. \cite{Shuman2016}, the downsampled signal was interpolated using the method of Pesenson \cite{Pesens2009} to reduce the prediction error as much as possible. In the present study, a symmetric structure was used; that is, simple upsampling followed by low-pass filtering was used as in the case of the original Laplacian pyramid for time/spatial domain signals \cite{Burt1983} since the purpose here was to compare pure downsampling and upsampling performances experimentally.

Two graph signals on the Minnesota Traffic Graph and a random sensor graph were decomposed. The signals are shown in Fig. \ref{fig:inputsignals}. That on the Minnesota Traffic Graph was decomposed into three levels, whereas that on the random sensor graph was decomposed into five levels. The graph-reduction method used is based on Kron reduction \cite{Dorfle2013} with edge sparsification. The low-pass filter used was $h(\la) = g(\la) = 1/(1+2\la)$ for all levels with a Chebyshev polynomial approximation \cite{Hammon2011,Shuman2011}.

The decomposed signals in each subband are shown scale-by-scale in Fig. \ref{fig:GLP_decomposition}. For the signals on the Minnesota Traffic Graph, a sharp transition between the upper and lower parts of the graph remained at all levels when a pyramid using (GD1) and (GU1) was applied. This indicates that the high-frequency graph components were not extracted into the corresponding high-frequency subbands. In contrast, pyramids created using the proposed sampling methods extracted different graph frequencies in different subbands.

Pyramids created using the proposed sampling methods can also be used for nonlinear approximation. For reconstruction, all coefficients in the coarsest subband and the $N_\text{kept}$ coefficients with the largest magnitude in the high-frequency subband are retained. The remaining coefficients are set to zero. The normalized errors $||\bm{f} - \bm{f}_\text{pred}||_2/||\bm{f}||_2$ given by the fraction of retained coefficients $N_\text{kept}/N$, where $\bm{f}_\text{pred}$ is the reconstructed graph signal after nonlinear approximation, are compared in Fig. \ref{fig:NLA_error}. The pyramids with spectral domain sampling outperform that based on the vertex domain approach. Overall, in this example, the index-based methods (GD2) and (GU2) perform better than the spectrum-based ones (GD3) and (GU3) in terms of decomposition quality and reconstruction error.

\section{Conclusions}\label{sec:conclusion}
The methods presented for sampling graph signals in the graph spectral domain inherit the expected spectral properties of classical sampled signals, such as bandwidth broadening by downsampling, unlike methods based on the conventional vertex graph sampling approach. Illustrative examples and potential applications demonstrate the validity of the proposed sampling methods as alternatives to the intuitive vertex graph sampling methods. Spectral domain sampling can be applied to both down- and upsampling of signals. Future work includes devising fast computation methods for spectral domain sampling, conducting experiments using real-world signals on (possibly large-scale) graphs, and designing multirate graph signal processing systems like wavelets \cite{Watana2018, Sakiya2018}.

\section*{Acknowledgment}
The author would like to thank the anonymous reviewers for their valuable comments and suggestions to improve the quality of the paper.


\end{document}